\documentclass[prb,twocolumn,superscriptaddress,amsmath,amssymb]{revtex4}

\renewcommand{\theequation}{\arabic{section}.\arabic{equation}}

\usepackage{dcolumn} 
\usepackage{epsfig}
\usepackage{subfigure}

\usepackage{theorem}
\theoremstyle{break}
\newtheorem{theorem}{Proposition}
\newtheorem{lemma}{Lemma}

\newcommand{\pdiff}[2]{\frac{\partial#1}{\partial#2}}

\newcommand{\Span}{\operatorname{span}}
\newcommand{\Hilb}{{\mathcal{H}}}
\newcommand{\Proj}{{\mathcal{P}}}
\newcommand{\Model}{{\mathcal{M}}}
\newcommand{\CalS}{{\mathcal{S}}}
\newcommand{\PhiFD}{\Phi^\text{FD}}
\newcommand{\Ind}{{\mathcal{I}}}
\newcommand{\rmd}{\mathrm{d}}

\newcommand{\eofproof}{\hspace{1em}$\diamondsuit$}
\newcommand{\braket}[1]{\langle{#1}\rangle}

\begin{document}

\title{Harmonic oscillator eigenfunction
  expansions, quantum dots, and effective interactions}

\date{\today}

\author{Simen Kvaal}
\email{simen.kvaal@cma.uio.no}
\affiliation{Centre of Mathematics for Applications, University of
  Oslo, N-0316 Oslo, Norway}

\begin{abstract}
  We give a thorough analysis of the convergence properties of the
  configuration-interaction method as applied to parabolic quantum
  dots among other systems, including \emph{a priori} error
  estimates. The method converges slowly in general, and in order to
  overcome this, we propose to use an effective two-body interaction
  well-known from nuclear physics. Through numerical experiments we
  demonstrate a significant increase in accuracy of the configuration
  interaction method.
\end{abstract}

\maketitle

\section{Introduction}
\setcounter{equation}{0}

The last two decades, an ever-increasing amount of research have been
dedicated to understanding the electronic structure of so-called quantum
dots\cite{Reimann2002}: semiconductor structures confining from
a few to several thousands electrons in spatial regions on the nanometre
scale. In such calculations, one typically seeks a few of the lowest
eigenenergies $E_k$ of the system Hamiltonian $H$ and their
corresponding eigenvectors $\psi_k$, i.e.,
\begin{equation} H\psi_k = E_k\psi_k, \quad k =
  1,\cdots,k_\text{max}. 
\end{equation} 
One of the most popular methods is the (full) configuration
interaction method (CI), where the many-body wave function is expanded
in a basis of eigenfunctions of the harmonic oscillator (HO), and then
necessarily truncated to give an approximation. In fact, the so-called
curse of dimensionality implies that the number of degrees of freedom
available per particle is severely limited. It is clear, that an
understanding of the properties of such basis expansions is very
important, as it is necessary for \emph{a priori} error estimates of
the calculations. Unfortunately, this is a neglected topic in the
physics literature.

In this article, we give a thorough analysis of the (full)
configuration interaction (CI) method using HO expansions applied to
parabolic quantum dots, and give practical convergence estimates. It
generalizes and refines the findings of a recent study of
one-dimensional systems,\cite{Kvaal2007} and is applicable to for
example nuclear systems\cite{HjorthJensen1995} and quantum chemical
calculations\cite{Helgaker2002} as well. We demonstrate the estimates
with calculations in the $d=2$ dimensional case for $N\leq 5$
electrons, paralleling computations in the
literature\cite{Maksym1990,Reimann2000,Bruce2000,Mikhailov2002,Rontani2006,Wensauer2004}.

The main results are however somewhat discouraging. The expansion coefficients
of typical eigenfunctions are shown to decay very slowly, limiting the
accuracy of \emph{any} practical method using HO basis functions. We
therefore propose to use an effective two-body interaction to
overcome, at least partially, the slow convergence rate. This is
routinely used in nuclear physics\cite{HjorthJensen1995,Navratil2000}
where the inter-particle forces are of a completely different, and
basically unknown, nature. For electronic systems, however, the
interaction is well-known and simpler to analyze, but effective
interactions of the present kind have not been applied, at least to
the author's knowledge.
The modified method is seen to have convergence rates of at least one
order of magnitude higher than the original CI method.
An important point here is that the complexity of the CI calculations
is not altered, as no extra non-zero matrix elements are
introduced. All one needs is a relatively simple one-time calculation
to produce the effective interaction matrix elements.

The HO eigenfunctions are popular for several reasons. Many quantum
systems, such as the quantum dot model considered here, are perturbed
harmonic oscillators \emph{per se}, so that the true eigenstates
should be perturbations of the HO states. Moreover, the HO has many
beautiful properties, such as complete separability of the
Hamiltonian, invariance under orthogonal coordinate changes, and thus
easily computed eigenfunctions, so that computing matrix elements of
relevant operators becomes relatively simple. The HO eigenfunctions
are defined on the \emph{whole} of $\mathbb{R}^d$ in which the
particles live, so that truncation of the domain is
unnecessary. Indeed, this is one of the main problems with methods
such as finite difference or finite element
methods.\cite{RamMohan2002} On the other hand, the HO eigenfunctions
are the \emph{only} basis functions with all these properties.

The article is organized as follows. In Sec.~\ref{sec:model} we
discuss the harmonic oscillator and the the parabolic quantum dot
model, including exact solutions for the $N=2$ case. In
Sec.~\ref{sec:series-new}, we give results for the approximation
properties of the Hermite functions in $n$ dimensions, and thus also
of many-body HO eigenfunctions.  By approximation properties, we mean
estimates on the error $\|\psi - P\psi\|$, where $\psi$ is any wave
function and $P$ projects onto a finite subspace of HO eigenfunctions,
i.e., the model space.  Here, $P\psi$ is in fact the best
approximation in the norm. The estimates will depend on analytic
properties of $\psi$, i.e., whether it is differentiable, and whether
it falls of sufficiently fast at infinity. To our knowledge, these
results are not previously published.

In Sec.~\ref{sec:ci}, we discuss the full configuration
interaction method, using the results obtained in
Sec.~\ref{sec:series-new} to obtain convergence estimates of the
method as function of the model space size. We also briefly discuss
the effective interaction utilized in the numerical calculations,
which are presented in Sec.~\ref{sec:numeric}. We conclude with a
discussion of the results, its consequences, and an outlook on further
directions of research in Sec.~\ref{sec:conclusion}.

We have also included an appendix with proofs of the formal
propositions in Sec.~\ref{sec:series-new}.

\section{The Harmonic Oscillator and parabolic quantum dots}
\setcounter{equation}{0}

\label{sec:model}

\subsection{The Harmonic Oscillator}
\label{sec:ho}

A spinless particle of mass $m$ in an isotropic harmonic potential has
Hamiltonian
\begin{equation} H_{\text{HO}} = -\frac{\hbar^2}{2m}\nabla^2 +
\frac{1}{2}m\omega^2\|\vec{r}\|^2, \end{equation} where $\vec{r}\in\mathbb{R}^d$
is the particle's coordinates. By choosing proper energy and length
units, i.e., $\hbar\omega$ and $\sqrt{\hbar/m\omega}$, respectively,
the Hamiltonian becomes
\begin{equation} H_\text{HO} = -\frac{1}{2}\nabla^2 + \frac{1}{2}\|\vec{r}\|^2. \end{equation}

$H_{\text{HO}}$ can be written as a sum over $d$ one-dimensional harmonic
oscillators, viz,
\begin{equation} H_{\text{HO}} = \sum_{k=1}^d \left(-\frac{1}{2}\pdiff{^2}{r_k^2} +
\frac{1}{2}r_k^2 \right),
\end{equation}
so that a complete specification of the HO eigenfunctions is given by
\begin{equation} \Phi_{\alpha_1,\alpha_2,\ldots,\alpha_d}(\vec{r}) =
\phi_{\alpha_1}(r_1)\phi_{\alpha_2}(r_2)\cdots\phi_{\alpha_d}(r_d), \label{eq:hermite} \end{equation}
where $\phi_{\alpha_i}(x)$, $\alpha_i=0,1,\ldots$ are one-dimensional HO
eigenfunctions, also called Hermite functions.  These are defined by
\begin{equation} \phi_n(x) = (2^n n! \pi^{1/2})^{-1/2} H_n(x)
  e^{-x^2/2}, \quad n = 0,1,\ldots, \label{eq:hermite1} \end{equation}
where the Hermite polynomials $H_n(x)$ are given by
\begin{equation} H_n(x) = (-1)^n e^{x^2} \pdiff{^n}{x^n} e^{-x^2}. \end{equation}
The Hermite polynomials also obey the recurrence formula
\begin{equation} H_{n+1}(x) = 2xH_n(x) - 2nH_{n-1}(x),
  \label{eq:recurrence} \end{equation}
with $H_0(x) = 1$ and $H_1(x) = 2x$. 
The Hermite polynomial $H_n(x)$ has $n$ zeroes, and the Gaussian
factor in $\phi_n(x)$ will eventually subvert the polynomial for large $|x|$. Thus,
qualitatively, the Hermite functions can be described as localized
oscillations with $n$ nodes and a Gaussian ``tail'' as $x$ approaches
$\pm\infty$. One can easily compute the quantum mechanical variance
\begin{equation} (\Delta x)^2 := \int_{-\infty}^\infty x^2\phi_n(x)^2 \rmd x = n + \frac{1}{2}, \end{equation}
showing that, loosely speaking, the width of the oscillatory region
increases as $(n+1/2)^{1/2}$. 

The functions $\Phi_{\alpha_1,\cdots,\alpha_d}$ defined in
Eqn.~(\ref{eq:hermite}) are called $d$-dimensional Hermite
functions. In the sequel, we will define $\alpha =
(\alpha_1,\cdots,\alpha_d)\in \Ind_d$ for a tuple of non-negative
integers, also called a multi-index; see Appendix
\ref{sec:multi-indices}. Using multi-indices, we may write
\begin{equation} \Phi_\alpha(\vec{r}) =
\left(2^{|\alpha|}\alpha!\pi^{d/2}\right)^{-1/2}
H_{\alpha_1}(r_1)\cdots H_{\alpha_d}(r_d)
e^{-\|\vec{r}\|^2/2}. \label{eq:d-hermite} \end{equation}

The eigenvalue of $\phi_n(x)$ is $n + 1/2$, so that the eigenvalue of
$\Phi_\alpha(\vec{r})$ is
\begin{equation} \epsilon_\alpha = \frac{d}{2} + |\alpha|, \end{equation}
i.e., a zero-point energy $d/2$ plus a non-negative integer. We denote
by $|\alpha|$ the \emph{shell number} of $\Phi_\alpha$, and the
eigenspace $\CalS_r(\mathbb{R}^d)$ corresponding to the eigenvalue
$d/2 + r$ a \emph{shell}. We define the
\emph{shell-truncated Hilbert space} $\Proj_R(\mathbb{R}^d) \subset
L^2(\mathbb{R}^d)$ as
\begin{equation} \Proj_R(\mathbb{R}^d) := \Span \big\{ \Phi_\alpha(\vec{r}) \; \big|
\; |\alpha| \leq R \big\} = \bigoplus_{r=0}^R
\CalS_r(\mathbb{R}^d), \end{equation} i.e., the subspace spanned by all Hermite
functions with shell number less than or equal to $R$, or,
equivalently, the direct sum of the shells up to and including
$R$. The $N$-body generalization of this space, to be discussed in
Section \ref{sec:many-body-wave-func}, is a very common model space
used in CI calculations.

Since the Hermite functions constitute an orthonormal basis for
$L^2(\mathbb{R}^d)$, $\Proj_R(\mathbb{R}^d) \rightarrow
L^2(\mathbb{R}^d)$, in the sense that for every $\psi\in
L^2(\mathbb{R}^d)$, $\lim_{R\rightarrow\infty} \|\psi-P\psi\| = 0$,
where $P$ is the orthogonal projector on
$\Proj_R(\mathbb{R}^d)$. Strictly speaking, we should use a symbol
like $P_R$ or even $P_R(\mathbb{R}^d)$ for the projector. However, $R$
and $d$ will always be clear from the context, so we are deliberately
sloppy to obtain a concise formulation. For the same reason, we will
sometimes simply write $\Proj$ or $\Proj_R$ for the space
$\Proj_R(\mathbb{R}^d)$.

An important fact is that since $H_\text{HO}$ is invariant under
orthogonal spatial transformations (i.e., such transformation conserve
energy), so is each individual shell space. Hence, each shell
$\CalS_r(\mathbb{R}^d)$, and also $\Proj_R(\mathbb{R}^d)$, is
independent of the spatial coordinates chosen.

For the case $d=1$ each shell $r$ is spanned by a single eigenfunction,
namely $\phi_r(x)$. For $d=2$, each shell $r$ has degeneracy $r+1$,
with eigenfunctions
\begin{equation} \Phi_{(s,r-s)}(\vec{r}) = \phi_s(r_1)\phi_{r-s}(r_2), \quad 0\leq s \leq r. \end{equation}

The usual HO eigenfunctions used to construct many-body wave functions
are not the Hermite functions $\Phi_{\alpha_1,\cdots,\alpha_d}$,
however, but rather those obtained by utilizing the spherical symmetry
of the HO. This gives a many-body basis diagonal in angular
momentum. For $d=2$ we obtain the so-called Fock-Darwin orbitals given
by
\begin{equation} \PhiFD_{n,m}(r,\theta) = \left[\frac{2n!}{(n+|m|)!}\right]^{1/2}\frac{e^{im\theta}}{\sqrt{2\pi}}
L_n^{|m|}(r^2)e^{-r^2/2}. \label{eq:fock-darwin} \end{equation}
Here, $n\geq 0$ is the nodal quantum number, counting the nodes of the
radial part, and $m$ is the azimuthal quantum number. 
The eigenvalues are
\begin{equation} \epsilon_{n,m} = 2n + |m| + 1. \end{equation}
Thus, $R = 2n + |m|$ is the shell number. By construction, the Fock-Darwin orbitals are
eigenfunctions of the angular momentum operator $L_z =
-i\partial/\partial_\theta$ with eigenvalue $m$.
Of course, we may write $\PhiFD_{n,m}$ as a
linear combination of the Hermite functions
$\Phi_{(s,R-s)}$, where $0\leq s \leq R = 2n + |m|$, and vice versa. 
The actual choice of form of eigenfunctions is immaterial, as long as
we may identify those belonging to a given shell.

\begin{figure*}
\includegraphics{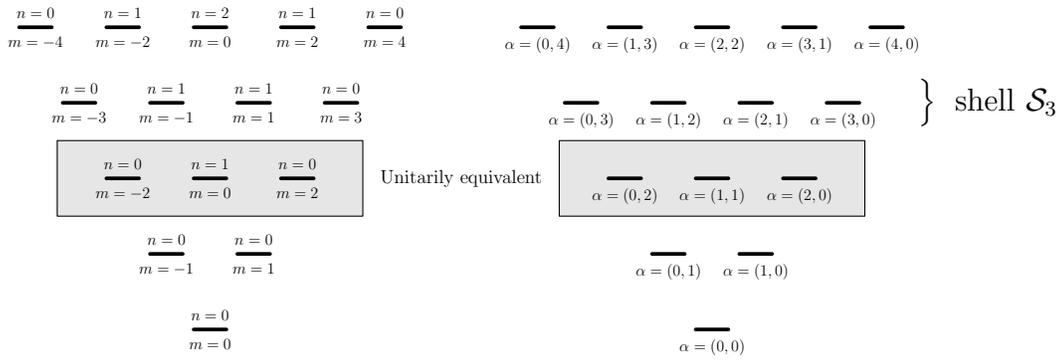}
\caption{\label{fig:shells}Illustration of $\Proj_{R=4}(\mathbb{R}^2)$: (Left)
  Fock-Darwin orbitals. (Right) Hermite functions. Basis functions
  with equal HO energy are shown at same line.}
\end{figure*}

The space $\Proj_{R=4}(\mathbb{R}^2)$ is illustrated in
Fig.~\ref{fig:shells} using both Hermite functions and Fock-Darwin
orbitals. 

\subsection{Parabolic quantum dots}
\label{sec:hamiltonian}

We consider $N$ electrons confined in a harmonic oscillator in $d$
dimensions. This is a very common model for a quantum dot. We comment,
that modelling the quantum dot geometry by a perturbed harmonic
oscillator is justified by self-consistent
calculations,\cite{Kumar1990,Macucci1997,Maksym1997} and is a widely
adopted
assumption.\cite{Maksym1990,Ezaki1997,Maksym1998,Tavernier2003,Wensauer2004,Rontani2006}

The Hamiltonian of the quantum dot is given by
\begin{equation} H := T + U, \label{eq:hamiltonian} \end{equation}
where $T$ is the many-body HO Hamiltonian, given by
\begin{equation}
  T = \sum_{k=1}^N H_\text{HO}(\vec{r}_k)
  \label{eq:many-body-ho}
\end{equation}
and $U$ is the inter-electron Coulomb interactions.  In dimensionless
units the interaction is given by,
\begin{equation} U := \sum_{i<j}^N C(i,j) = \sum_{i<j}^N \frac{\lambda}{\|\vec{r}_i - \vec{r}_j\|}.
\end{equation}
The $N$ electrons
have coordinates $\vec{r}_k$, and the parameter $\lambda$ measures the strength of the interaction over
the confinement of the HO, viz,
\begin{equation} \lambda := \frac{1}{\hbar\omega}\left(\frac{e^2}{4\pi\epsilon_0
  \epsilon}\right),\end{equation} where we recall that $\sqrt{\hbar/m\omega}$ is
the length unit. Typical values for GaAs semiconductors are close to
$\lambda = 2$, see for example
Ref.~\onlinecite{Tavernier2003}. Increasing the trap size leads to a
larger $\lambda$, and the quantum dot then approaches the classical
regime.\cite{Reimann2002}

\subsection{Exact solution for two electrons}
\label{sec:bad-example}

Before we discuss the approximation properties of the Hermite
functions, it is instructive to consider the very simplest example of
a two-electron parabolic quantum dot and the properties of the
eigenfunctions, since this case admits analytical solutions for
special values of $\lambda$ and is otherwise well
understood\cite{Taut1993,Taut1994,Pfannkuche1993}. Here, we consider
$d=2$ dimensions only, but the $d=3$ case is similar. We note, that
for $N=2$ it is enough to study the spatial wave function, since it
must be either symmetric (for the singlet $S=0$ spin state) or
anti-symmetric (for the triplet $S=1$ spin states). The Hamiltonian
\eqref{eq:hamiltonian} becomes
\begin{equation} H = -\frac{1}{2}(\nabla_1^2 + \nabla_2^2) + \frac{1}{2}(r_1^2 +
r_2^2) + \frac{\lambda}{r_{12}}, \label{eq:two-electron-ham} \end{equation}
 where $r_{12} = \| \vec{r}_1 -
\vec{r}_2 \|$ and $r_j = \|\vec{r}_j\|$. Introduce a set of scaled centre of mass
coordinates given by $\vec{R}=(\vec{r}_1+\vec{r}_2)/\sqrt{2}$ and $\vec{r} = (\vec{r}_1 - \vec{r}_2)/\sqrt{2}$.
This coordinate change is orthogonal and symmetric in $\mathbb{R}^4$.
This leads to the separable Hamiltonian
\begin{eqnarray*} H &=& -\frac{1}{2}(\nabla^2_r + \nabla^2_R) + \frac{1}{2}(\|\vec{r}\|^2 +
\|\vec{R}\|^2) + \frac{\lambda}{\sqrt{2}\|\vec{r}\|} \\ &=&
H_\text{HO}(\vec{R}) + H_\text{rel}(\vec{r}). \end{eqnarray*}
A complete set of eigenfunctions of $H$ can now be written on product form, viz,
\begin{equation} \Psi(\vec{R},\vec{r}) = \PhiFD_{n',m'}(\vec{R})\psi(\vec{r}). \end{equation}
The relative coordinate wave function $\psi(\vec{r})$ is an
eigenfunction of the relative coordinate Hamiltonian given by
\begin{equation} H_\text{rel} = -\frac{1}{2}\nabla_r^2 + \frac{1}{2}r^2 +
\frac{\lambda}{\sqrt{2}r}, \label{eq:H_r}
\end{equation} where $r = \|\vec{r}\|$. This 
Hamiltonian can be further separated using polar coordinates, yielding
eigenfunctions on the form
\begin{equation} \psi_{m,n}(r,\theta) = \frac{e^{im\theta}}{\sqrt{2\pi}}
u_{n,m}(r), \end{equation}
where $|m|\geq 0$ is an integer and $u_{n,m}$ is an eigenfunction of the radial
Hamiltonian given by
\begin{equation} H_r = -\frac{1}{2r}\pdiff{}{r}r\pdiff{}{r} +
\frac{|m|^2}{2r^2} + \frac{1}{2}r^2 + \frac{\lambda}{\sqrt{2}r}. \end{equation}
By convention, $n$ counts the nodes away from $r=0$ of
$u_{n,m}(r)$. Moreover, odd (even) $m$ gives anti-symmetric
(symmetric) wave functions $\Psi(\vec{r}_1,\vec{r}_2)$. For any given
$|m|$, it is quite easy to deduce that the special value $\lambda =
\sqrt{2|m| + 1}$ yields the eigenfunction
\begin{equation} u_{0,m} = D r^{|m|}(a + r)e^{-r^2/2}, \end{equation}
where $D$ and $a$ are constants. The corresponding eigenvalue of $H_r$
is $E_r = |m|+2$, and $E = 2n' + |m'| + 1 + E_r$. Thus, the ground
state (having $m=m'=0$, $n=n'=0$) for $\lambda=1$ is given by
\begin{eqnarray*} \Psi_0(\vec{R},\vec{r}) &=& D (r +
  a)e^{-(r^2+R^2)/2} \\
&=& \frac{D}{\sqrt{2}}(r_{12} + \sqrt{2}a) e^{-(r_1^2+r_2^2)/2},
 \end{eqnarray*}
with $D$ being a (new) normalization constant.

Observe that this function has a cusp at $r=0$, i.e., at the origin
$x=y=0$ (where we have introduced Cartesian coordinates
$\vec{r}=(x,y)$ for the relative coordinate). Indeed, the partial
derivatives $\partial_x\psi_{0,0}$ and $\partial_y\psi_{0,0}$ are not
continuous there, and $\Psi_0$ has no partial derivatives (in the
distributional sense, see Appendix \ref{sec:sobolev}) of second
order.  The cusp stems from the famous ``cusp condition'' which in
simple terms states that, for a non-vanishing wave function at
$r_{12}=0$, the Coulomb divergence must be compensated by a similar
divergence in the Laplacian.\cite{Kato1957,HoffmannOstenhof1992} This
is only possible if the wave function has a cusp.

On the other hand, the non-smooth function
$\Psi_{0}(\vec{R},\vec{r})$ is to be expanded in the HO
eigenfunctions, e.g., Fock-Darwin orbitals. (Recall, that the
particular representation for the HO eigenfunctions are immaterial --
also whether we use lab coordinates $\vec{r}_{1,2}$ or centre-of-mass
coordinates $\vec{R}$ and $\vec{r}$, since the coordinate change is
orthogonal.) For $m=0$, we have
\begin{equation} \PhiFD_{n,0}(r) = \sqrt{\frac{2}{\pi}} L_n(r^2)e^{-r^2/2}, \end{equation}
using the fact that these are independent of $\theta$.
Thus,
\begin{equation} \Psi_{0}(\vec{r}) = \PhiFD_{0,0}(R)u_{0,0}(r) = \PhiFD_{0,0}(R)\sum_{n=0}^\infty c_n
  \PhiFD_{n,0}(r), \label{eq:expansion} \end{equation} The functions
$\PhiFD_{n,0}(r)$ are very smooth, as is seen by noting that
$L_n(r^2)=L_n(x^2+y^2)$ is a polynomial in $x$ and $y$, while
$u_{0,0}(r) = u_{0,0}(\sqrt{x^2+y^2})$, so Eqn.~\eqref{eq:expansion}
is basically approximating a square root with a polynomial.

Consider then a truncated expansion $\Psi_{0,R} \in \Proj_R(\mathbb{R}^2)$, such as
the one obtained with the CI or coupled cluster method.\cite{Bartlett2007} In general,
this is different from $P_R\Psi_0$, which is the best approximation of
the wave function in $\Proj_R(\mathbb{R}^2)$. In any case, this expansion, consisting
of the $R+1$ terms like those of Eqn.~\eqref{eq:expansion} is a very
smooth function. Therefore, the cusp at $r=0$ cannot be well
approximated.

In Section \ref{sec:2d-app}, we will show that the smoothness
properties of the wave function $\Psi$ is equivalent to a certain
decay rate of the coefficients $c_n$ in Eqn.~\eqref{eq:expansion} as $n\rightarrow\infty$. In this
case, we will show that 
\begin{equation} \sum_{n=0}^\infty n^k|c_n|^2 < +\infty,  \end{equation}
so that
\begin{equation} |c_n| = o(n^{-(k + 1 + \epsilon)/2}). \label{eq:order-claim} \end{equation} Here, $k$ is
the number of times $\Psi$ may be differentiated weakly, i.e.,
$\Psi\in H^k(\mathbb{R}^2)$, and $\epsilon\in[0,1)$ is a constant. For the function $\Psi_0$ we have
$k=1$. This kind of estimate directly tells us that an approximation
using only a few HO eigenfunctions necessarily will give an error
depending directly on the smoothness $k$.

We comment, that for higher $|m|$ the eigenstates will still have
cusps, albeit in the higher
derivatives.\cite{HoffmannOstenhof1992} Indeed, we have weak
derivatives of order $|m|+1$, as can easily be deduced by operating on
$\psi_{0,m}$ with $\partial_x$ and $\partial_y$. Moreover, recall that
$|m|=1$ is the $S=1$ ground state, which then will have coefficients
decaying faster than the $S=0$ ground state. Moreover, there will be
excited states, i.e., states with $|m|>1$, that also have more quickly
decaying coefficients $|c_n|$. This will be demonstrated numerically
in Sec.~\ref{sec:numeric}. 

In fact, Hoffmann-Ostenhof \emph{et al.}\cite{HoffmannOstenhof1992}
have shown that near $r_{12}=0$, for arbitrary $\lambda$ \emph{any}
local solution $\Psi$ of $(H-E)\Psi=0$ has the form
\begin{equation} \Psi(\xi) = \|\xi\|^m P\left(\frac{\xi}{\|\xi\|}\right)(1 + a\|\xi\|) + O(\|\xi\|^{m+1}), \end{equation}
where $\xi = (\vec{r}_1,\vec{r_2})\in\mathbb{R}^4$, and where $P$,
$\deg(P) = m$, is a hyper-spherical harmonic (on $S^3$), and where $a$
is a constant. This also generalizes to arbitrary $N$,
cf.~Sec.~\ref{sec:analytic-props}. From this representation, it is manifest,
that $\Psi\in H^{m+1}(\mathbb{R}^4)$, i.e., $\Psi$ has weak
derivatives of order $m+1$. We discuss these results further in
Sec.~\ref{sec:analytic-props}.

\section{Approximation properties of Hermite series}
\setcounter{equation}{0}
\label{sec:series-new}

\subsection{Hermite functions in one dimension}
\label{sec:hermite-analysis}

In this section, we consider some formal mathematical propositions
whose proofs are given in Appendix \ref{app:proofs}, and discuss
their importance for expansions in HO basis functions. 

The first proposition considers the one-dimensional case, and the
second considers general, multidimensional expansions.  The latter
result has to the author's knowledge not been published
previously. The treatment for one-dimensional Hermite functions is
similar, but not equivalent to, that given by Boyd\cite{Boyd1984} and
Hille.\cite{Hille1939} 

We stress that the results are valid for \emph{any} given
wavefunction -- not only eigenfunctions of quantum dot
Hamiltonians -- assuming only that the wavefunction decays
exponentially as $|x|\rightarrow\infty$. In Appendix
\ref{app:proofs}, more general conditions are also considered.

The results are stated in terms of \emph{weak} differentiability of the
wavefunction, which is a generalization of the classical
notion of a derivative. The space $H^k(\mathbb{R})\subset L^2(\mathbb{R})$ is
roughly defined as the (square integrable) functions $\psi(x)$ having
$k$ (square integrable) derivatives $\partial^m_x\psi(x)$, $0\leq m
\leq k$. Correspondingly, the space $H^k(\mathbb{R}^n)\subset
L^2(\mathbb{R}^n)$ consists of the functions whose partial derivatives
of total order $\leq k$ are square integrable. For wavefunctions of
electronic systems, it turns out that $k$ times continuous
differentiability implies $k+1$ times weak
differentiability\cite{HoffmannOstenhof1992}. The order $k$ of
differentiability is not always known, but an upper or lower bound can
often be found through analysis. It is however important, that the
Coulomb singularity implies that $k$ is finite.

For the one-dimensional case, we have the following proposition:
\begin{theorem}[Approximation in one dimension]\label{prop:approx1b}
Let $k \geq 0$ be a given integer. Let $\psi \in L^2(\mathbb{R})$ be
exponentially decaying as $|x|\rightarrow\infty$ and 
given by
\begin{equation} \psi(x) = \sum_{n=0}^\infty c_n \phi_n(x), \end{equation}
where $\phi_n(x)$ is given by Eqn.~(\ref{eq:hermite1}).
Then $\psi\in H^k(\mathbb{R})$ if and only if
\begin{equation} \sum_{n=0}^\infty n^k |c_n|^2 < \infty. \end{equation}
\end{theorem}

We notice that the latter implies that 
\begin{equation} |c_n| =
  o(n^{-(k+1)/2}), 
\end{equation} 
which shows that the more $\psi(x)$ can be differentiated, the faster
the coefficients will fall off as $n\rightarrow\infty$.  Moreover, let
$\psi_R = P_R\psi = \sum_{n=0}^R c_n\phi_n$. Then
\begin{equation} \| \psi - \psi_R \| = \left( \sum_{n=R+1}^\infty |c_n|^2
\right)^{1/2}, \end{equation}
which gives an estimate of how well a finite basis of Hermite
functions will approximate $\psi(x)$ in the norm. We already notice,
that for low $k=2$, which is typical, the coefficients fall off as
$o(n^{-3/2})$, which is rather slowly.

In the general $n$-dimensional case, the wavefunction $\psi\in
L^2(\mathbb{R}^n)$ has an expansion in the $n$-dimensional Hermite
functions $\Phi_\alpha(x)$, $\alpha\in \Ind_n$ given by
\begin{eqnarray*}
  \psi(x) &=& \sum_\alpha c_\alpha \Phi_\alpha(x) \\ &=&
  \sum_{\alpha_1\cdots\alpha_n} c_{\alpha_1\cdots\alpha_n}\phi_{\alpha_1}(x_1)\cdots\phi_{\alpha_n}(x_n).
\end{eqnarray*}
In order to obtain useful estimates on the error, we need to define the
\emph{shell-weight} $p(R)$ by the overlap of $\psi(x)$ with the single
shell
$\CalS_R$, i.e.,
\begin{equation}
  p(R) = \|P(\CalS_R)\psi\|^2 = \sum_{\alpha,|\alpha|=R} |c_\alpha|^2,
\end{equation}
where $P(\CalS_R)$ is the projection onto the shell. Thus,
\begin{equation}
  \|\psi\|^2 = \sum_{R=0}^\infty p(R).
\end{equation}
For the one-dimensional case, we of course have $p(R)=|c_R|^2$.
\begin{theorem}[Approximation in $n$ dimensions]\label{prop:approx-general-n}
Let $\psi \in L^2(\mathbb{R}^n)$ be exponentially decaying  as
$\|x\|\rightarrow\infty$ and given by
\begin{equation} \psi(x) = \sum_\alpha c_\alpha \Phi_\alpha(x). \end{equation}
Then $\psi\in H^k(\mathbb{R}^n)$  if and only if
\begin{equation} \sum_\alpha |\alpha|^k |c_\alpha|^2 = \sum_{r=0}^\infty r^k p(r)  <
+\infty. \label{eq:multisum} \end{equation}
\end{theorem}
Again, we notice that the latter implies that 
\begin{equation} p(r) =
  o(r^{-(k+1)}). \label{eq:order-nd} \end{equation} Moreover,
for the shell-truncated Hilbert space $\Proj_R$, the approximation
error is given by
\begin{equation} \| (1-P)\psi \| = \left( \sum_{r=R+1}^\infty
  p(r)\right)^{1/2}. \label{eq:approx-error-nd} \end{equation}

In applications, we often observe a decay of non-integral order, i.e.,
there exists an $\epsilon\in[0,1)$ such that we observe
\begin{equation} p(r) =
  o(r^{-(k+1+\epsilon)}). \label{eq:order-nd-eps} \end{equation} This
does not, of course, contradict the results. To see this, we observe
that if $\psi\in H^k(\mathbb{R}^n)$ but $\psi\notin
H^{k+1}(\mathbb{R}^n)$, then $p(r)$ must decay at least as fast as
$o(r^{-(k+1)})$ but not as fast as $o(r^{-k+2})$. Thus, the actual
decay exponent can be anything inside the interval $[k+1,k+2)$.

Consider also the case where $\psi\in H^k(\mathbb{R}^n)$ for every
$k$, i.e., we can differentiate it (weakly) as many times we
like. Then $p(r)$ decays faster than $r^{-(k+1)}$, for any $k\geq 0$,
giving so-called exponential convergence of the Hermite series. Hence,
functions that are best approximated by Hermite series are rapidly
decaying \emph{and} very smooth functions $\psi$. This would be the
case for the quantum dot eigenfunctions if the inter-particle
interactions were non-singular.

\subsection{Many-body wave functions}
\label{sec:many-body-wave-func}

We now discuss $N$-body eigenfunctions of the HO in $d$ dimensions,
including spin, showing that we may identify the expansion of a
such with $2^N$ expansions in Hermite functions in
$n=Nd$ dimensions, i.e., $2^N$ expansions in HO eigenfunctions of
imagined spinless particles in $n=Nd$ dimensions. Each expansion
corresponds to a different spin configuration.

Each particle $k = 1,\ldots,N$ has both spatial degrees of freedom
$\vec{r}_k\in\mathbb{R}^d$ and a spin coordinate $\tau_k \in \{ \pm 1
\} $, corresponding to the $z$-projection $S_z = \pm\frac{\hbar}{2}$
of the electron spin. The configuration space can thus be taken as two
copies $X$ of $\mathbb{R}^d$; one for each spin value, i.e., $X =
\mathbb{R}^d \times \{ \pm 1 \}$ and $x_k = (\vec{r}_k,\tau_k)\in X$
are the coordinates of particle $k$.

For a single particle with spin, the Hilbert space is now $L^2(X)$,
with basis functions given by
\begin{equation}
  \hat{\Phi}_i(x) = \Phi_\alpha(\vec{r})\chi_\sigma(\tau),
\end{equation}
where $i=i(\alpha,\sigma)$ is a new, generic index, and where
$\chi_\sigma$ is a basis function for the spinor space
$\mathbb{C}^2$. 

Ignoring the Pauli principle for the moment, the $N$-body Hilbert
space is now given by
\begin{equation} \Hilb(N) = L^2(X)^N \equiv L^2(\mathbb{R}^{Nd})\otimes
(\mathbb{C}^2)^{N}, \end{equation} i.e., each wavefunction $\psi\in
\mathcal{H}(N)$ is equivalent to $2^N$ spin-component
functions $\psi^{(\sigma)}\in L^2(\mathbb{R}^{Nd})$, $\sigma =
(\sigma_1,\cdots,\sigma_N) \in \{
\pm 1 \}^N$. We have
\begin{equation}
 \psi(x_1,\cdots,x_N) = \sum_\sigma
\psi^{(\sigma)}(\xi)\chi_\sigma(\tau), \quad \xi\equiv(
\vec{r}_1,\cdots,\vec{r}_N) ,
\end{equation}
 where
$\tau = (\tau_1,\cdots,\tau_N)$, and where $\chi_\sigma(\tau) =
\delta_{\sigma,\tau}$ are basis functions for the $N$-spinor space
$\mathbb{C}^{2^N}$, being eigenfunctions for $S_z$, i.e.,
corresponding to a given configuration of the $N$ spins.

The $\sigma$'th component function $\psi^{(\sigma)}(\xi)\in
L^2(\mathbb{R}^{Nd})$ is a function of $Nd$ variables, and by
considering the $Nd$-dimensional HO as the sum of $N$ HOs in $d$
dimensions, it is easy to see that a basis for the $L^2(\mathbb{R}^{Nd})$ is
given by the functions
\begin{equation} \Phi_\beta(\xi) \equiv \Phi_{\alpha^1}(\vec{r}_1)\cdots
\Phi_{\alpha^N}(\vec{r}_N) \end{equation}
where $\xi=(\vec{r}_1,\cdots,\vec{r}_N)$, and where
$\beta=(\alpha^1,\cdots,\alpha^N)$ is an $Nd$-component
multi-index. Correspondingly, a basis for the complete space
$\Hilb(N)=L^2(X)^N$ is given by the functions
\begin{equation} \hat{\Phi}_{i_1\cdots i_N}(\xi,\tau) \equiv \Phi_{\alpha^1}(\vec{r}_1)\cdots
\Phi_{\alpha^N}(\vec{r}_N)\chi_\sigma(\tau), \end{equation}
where $i_k = i(\alpha^k,\sigma_k)$. Notice, that the HO energy and
hence the shell number $|\beta|$ only
depends on $\beta=(\alpha^1,\cdots,\alpha^N)$.

The functions $\psi^{(\sigma)}$ may be expanded in the functions $\Phi_\beta$,
i.e.,
\begin{eqnarray*}
  \psi^{(\sigma)}(\xi) &=&  \sum_\beta c^{(\sigma)}_\beta
  \Phi_\beta(\xi) \\
 &=& \sum_{\alpha^1\cdots\alpha^N}
 c^{(\sigma)}_{\alpha^1\cdots\alpha^N} \Phi_{\alpha^1}(\vec{r}_1)\cdots\Phi_{\alpha^N}(\vec{r}_N),
\end{eqnarray*}
and we define the $\sigma$'th shell weight $p^{(\sigma)}(r)$ as
before, i.e.,
\begin{equation}
  p^{(\sigma)}(r) \equiv \sum_{|\beta| = r} |c^{(\sigma)}_\beta|^2.
\end{equation}
We may then apply the analysis from Section \ref{sec:hermite-analysis}
to each of the spin component functions, and note that the
\emph{total} shell-weight is
\begin{equation}
  p(r) \equiv \sum_{i_1\cdots i_N} \langle
  \hat{\Phi}_{i_1\cdots i_N}, \psi\rangle \delta_{|\beta|,r}= \sum_{\sigma} p^{(\sigma)}(r)
\end{equation}
since the shell number $|\beta| = \sum|\alpha_k|$ does not depend on
the spin configuration of the basis function.

Including the Pauli principle to accommodate proper wave-function
symmetry does not change these considerations. The basis functions
$\hat{\Phi}_{i_1\cdots i_N}$ are anti-symmetrized to become Slater
determinants ${\Psi}_{i_1\cdots i_N}$ (see for example
Ref.~\onlinecite{Raimes1972} for details), which is equivalent to
consider the projection $\Hilb_\text{AS}(N)=P_\text{AS}\Hilb(N)$ of
the unsymmetrized space onto the antisymmetric subspace. Moreover, the
projections $P_R$ and $P_\text{AS}$ commute, so that
the shell-truncated
space is given by
\begin{equation}
  \Proj_{\text{AS},R} = \Span \left\{ \Psi_{i_1,\cdots,i_N} \; : \; i_1=i <
    \cdots < i_N, \; \sum_k |\alpha^k| \leq R \right\} , 
\end{equation}
which is precisely the computational basis used in many CI
calculations. (See however also the discussion in Section
\ref{sec:ci}.) 
We stress, that, $\Proj_{\text{AS},R}$ is independent of the actual
one-body HO eigenfunctions used. 
The shell-weight of $\psi\in\Hilb_\text{AS}(X)$ is now
given by
\begin{equation}
  p(r) = \sum_{i_1\cdots i_N} \langle {\Psi}_{i_1\cdots i_N},
  \psi\rangle \delta_{|\beta(i_1\cdots i_N)|,r},
\end{equation}
and 
\begin{equation}
  \| P_R \psi \|^2 = \sum_{r=0}^R p(r).
\end{equation}

As should be clear now, studying approximation of Hermite functions in arbitrary
dimensions automatically gives the corresponding many-body
HO approximation properties, since the many-body eigenfunctions can be seen
as $2^N$ component functions, and since the shell-truncated Hilbert
space transfers to a many-body setting in a natural way.

\subsection{Two electrons revisited}
\label{sec:2d-app}

We return to the exact solutions of the two-electron quantum dot
considered in Sec.~\ref{sec:bad-example}. Recall, that the wave
functions were on the form
\begin{equation} \psi(r,\theta) = e^{im\theta}f(r), \end{equation}
where $f(r)$ decayed exponentially fast as $r\rightarrow\infty$.
Assume now, that
$\psi\in H^k(\mathbb{R}^2)$, i.e., that all partial derivatives of
$\psi$ of order $k$ exists in the weak sense, viz,
\begin{equation} \partial_x^j\partial_y^{k-j}\psi\in L^2(\mathbb{R}^2), \quad 0\leq
j \leq k, \end{equation}
where $x = r\cos(\theta)$ and $y=r\sin(\theta)$.
Then, by Lemma~\ref{thm:exp-decay} in the Appendix,
$(a^\dag_x)^j(a^\dag_y)^{k-j}\psi\in L^2(\mathbb{R}^2)$ for $0\leq j
\leq k$ as well.

The function $\psi(r,\theta)$ was expanded in Fock-Darwin orbitals, viz,
\begin{equation} \psi(r,\theta) = \sum_{n=0}^\infty c_n
\PhiFD_{n,m}(r,\theta). \end{equation}

Recall, that the shell number $N$ for $\PhiFD_{n,m}$ was given by
$N=2n+|m|$. Thus, the shell-weight $p(N)$ is in this case simply
\begin{equation} p(N) = |c_{(N-|m|)/2}|^2, \quad N\geq |m|, \end{equation}
and $p(N)=0$ otherwise. From Prop.~\ref{prop:approx-general-n}, we
have
\begin{equation} \sum_{N=|m|}^\infty N^k p(N) < +\infty, \end{equation}
which yields
\begin{equation} |c_n| = o(n^{-(k+1 + \epsilon)/2}), \quad 0\leq\epsilon<1, \end{equation}
as claimed in Sec.~\ref{sec:bad-example}.

\subsection{Smoothness properties of many-electron wave functions}
\label{sec:analytic-props}

Let us mention some results, mainly due to
Hoffmann-Ostenhof \emph{et
  al.},\cite{HoffmannOstenhof1992,HoffmannOstenhof1994} concerning
smoothness of many-electron wave functions. Strictly speaking, their
results are valid only in $d=3$ spatial dimensions, since the Coulomb
interaction in $d=2$ dimensions fails to be a Kato potential, the
definition of which is quite subtle and out of the scope for this
article\cite{HoffmannOstenhof1994}. On the other hand, it is
reasonable to assume that the results will still hold true, 
since the analytical results of the $N=2$ case is very similar in the
$d=2$ and $d=3$ cases: The eigenfunctions decay exponentially with the
same cusp singularities at the origin.\cite{Taut1993,Taut1994}

Consider the Schr\"odinger equation $(H-E)\psi(\xi) = 0$, where
$\xi=(\xi_1,\cdots,\xi_{Nd}) = (\vec{r}_1,\cdots,\vec{r}_N) \in
\mathbb{R}^{Nd}$, and where $\psi(\xi)$ is only assumed to be a
solution locally. (A proper solution is of course also a local solution.)
Recall, that $\psi$ has $2^N$ spin-components $\psi^{(\sigma)}$. Define a coalesce point
$\xi_\text{CP}$ as a point where at least two particles coincide,
i.e., $\vec{r}_j=\vec{r}_\ell$, $j\neq \ell$. Away from the set of such
points, $\psi^{(\sigma)}(\xi)$ is real analytic, since the interaction is real
analytic there. Near a $\xi_\text{CP}$, the wave function has the form
\begin{equation} \psi^{(\sigma)}(\xi + \xi_\text{CP}) = r^k P(\xi/r)( 1 + a r ) + O(r^{k+1}), \end{equation}
where $r = \|\xi\|$, $P$ is a hyper-spherical harmonic (on the sphere $S^{Nd-1}$) of
degree $k = k(\xi_\text{CP})$, and where $a$ is a constant. It is immediately clear, that
$\psi^{(\sigma)}(\xi)$ is $k+1$ times weakly differentiable in a neighborhood of
$\xi_\text{CP}$. However, at $K$-electron coalesce points, i.e., at
points $\xi_\text{CP}$ where $K$ different electrons coincide, the
integer $k$ may differ. Using exponential decay
of a proper eigenfunction, we have $\psi^{(\sigma)}\in
H^{\min(k)+1}(\mathbb{R}^{Nd})$. Hoffmann-Ostenhof \emph{et al.} also
showed, that symmetry restrictions on the spin-components due to the
Pauli principle induces an increasing degree $k$ of the hyper-spherical
harmonic $P$, generating even higher order of smoothness. A general
feature, is that the smoothness increases with the number of particles.

However, their results in this direction are not general enough to
ascertain the \emph{minimum} of the values for $k$ for a given wave
function, although we feel rather sure that such an analysis is
possible. Suffice it to say, that the results are clearly visible in
the numerical calculations in Sec.~\ref{sec:numeric}.

Another interesting direction of research has been undertaken by
Yserentant,\cite{Yserentant2004} who showed that there are some very
high order mixed partial derivatives at coalesce points. It seems
unclear, though, if this can be exploited to improve the CI
calculations further.

\section{The configuration interaction method}
\setcounter{equation}{0}

\label{sec:ci}

\subsection{Convergence analysis using HO eigenfunction basis}
\label{sec:ci-convergence}

The basic problem is to determine a few eigenvalues
and eigenfunctions of the Hamiltonian $H$ in Eqn.~(\ref{eq:hamiltonian}), i.e.,
\begin{equation} H\psi_k = E_k\psi_k, \quad k =
  1,\cdots,k_\text{max}. \end{equation} The CI method consists of
approximating eigenvalues of $H$ with those obtained by projecting the
problem onto a finite-dimensional subspace $\Hilb_h \subset
\Hilb(N)$. As such, it is an example of the Ritz-Galerkin variational
method.\cite{Babuska1987,Babuska1996} We comment, that the convergence
of the Ritz-Galerkin method is \emph{not} simply a consequence of the
completeness of the basis functions.\cite{Babuska1996} We will analyze
the CI method when the model space is given by
\begin{eqnarray*} \Hilb_h &=& P_R\Hilb_\text{AS}(N) = \Proj_{R}(N) \\ &=& \Span \left\{
\Psi_{i_1,\cdots,i_N} \; : \; \sum_k |\alpha^k| \leq R \right\} , \end{eqnarray*}
used in Refs.~\onlinecite{Mikhailov2002,Tavernier2006}, for example,
although other spaces also are common. (We drop the subscript ``AS''
from now on.) The space
\begin{equation} \Model_R(N) := \Span \left\{ \Psi_{i_1,\cdots,i_N} \;
    : \; \max_k |\alpha^k| \leq R \right\}, \end{equation} i.e., a cut
in the \emph{single-particle} shell numbers (or energy) instead of the
global shell number (or energy) is also common.\cite{Reimann2000,Rontani2006} For
obvious reasons, $\Proj_R(N)$ is often referred to as an ``energy cut
space'', while $\Model_R(N)$ is referred to as a ``direct product
space''. 

As in Sec.~\ref{sec:hermite-analysis}, $P_R$ is the
orthogonal projector onto the model space $\Proj_R(N)$. We also define
$Q_R = 1-P_R$ as the projector onto the excluded space $\Proj_R(N)^\bot$.
The discrete eigenvalue problem is then
\begin{equation} (P_RHP_R)\psi_{h,k} = E_{h,k}\psi_{h,k}, \quad k =
1,\cdots,k_\text{max}. \end{equation}

The CI method becomes, in principle, exact as $R \rightarrow \infty$.
Indeed, a widely-used name for the CI method is ``exact
diagonalization,'' being somewhat a misnomer as only a very limited
number of degrees of freedom per particle is achievable.

It is clear that
\begin{equation} \Proj_{R}(N) \subset \Model_R(N) \subset
  \Proj_{NR}(N), \label{eq:subspaces} \end{equation} so that studying
the convergence in terms of $\Proj_{R}(N)$ is sufficient. In our
numerical experiments we therefore focus on the energy cut model space. A
comparison between the convergence of the two spaces is, on the other
hand, an interesting topic for future research.

Using the results in Refs.~\onlinecite{Babuska1989,Babuska1996} for
non-degenerate eigenvalues for simplicity, we obtain an estimate for
the error in the numerical eigenvalue $E_h$ as
\begin{equation} E_h - E \leq [1 + \nu(R)](1 +
  K\lambda)\braket{\psi,Q_R T\psi},
  \label{eq:CI-eigenvalue-error} \end{equation}
where $K$ is a constant, and where $\nu(R)\rightarrow 0$ as
$R\rightarrow \infty$. Using $T\Phi_\beta = (Nd/2 + |\beta|)\Phi_\beta$ and
Eqn.~(\ref{eq:multisum}), we obtain
\begin{equation} \braket{\psi,Q_RT\psi} = \sum_{r=R+1}^\infty \left(\frac{Nd}{2} +
r\right) p(r). \end{equation}
Assume now, that $\psi^{(\sigma)}\in H^k(\mathbb{R}^{Nd})$ for all $\sigma$, so
that according to Proposition~\ref{prop:approx-general-n}, we will have
\begin{equation} \sum_{r=0}^\infty r^k p(r) < +\infty \end{equation}
implying that $rp(r) = o(r^{-k})$. We
then obtain, for $k>1$,
\begin{equation} \braket{\psi,(1-P_R)T\psi} = o(R^{-(k-1)}) + o(R^{-k}). \end{equation} For
$k=1$ (which is the worst case), we merely obtain convergence,
$\braket{\psi,(1-P_R)T\psi} \rightarrow 0$ as $R\rightarrow\infty$.
We assume, that $R$ is sufficiently large, so that the $o(R^{-k})$
term can be neglected.

Again, we may observe a slight deviation from the decay, and we expect
to observe eigenvalue errors on the form
\begin{equation} E_h - E \sim (1 + K\lambda) R^{-(k - 1 + \epsilon)}, \label{eq:eigenvalue-estimate} \end{equation}
where $0\leq \epsilon < 1$.

As for the eigenvector error $\|\psi_h - \psi\|$ (recall that $\psi_h
\neq P_R\psi$), we mention that
\begin{equation} \|\psi_h - \psi\| \leq [1 +
\eta(R)]\left[(1+K\lambda)\braket{\psi,(1-P_R)T\psi}\right]^{1/2}, \end{equation}
where $\eta(R)\rightarrow 0$ as $R\rightarrow \infty$.

\subsection{Effective interaction scheme}
\label{sec:effint}

Effective interactions have a long tradition in nuclear physics, where
the bare nuclear interaction is basically unknown and highly singular,
and where it must be renormalized and fitted to experimental
data.\cite{HjorthJensen1995} In quantum chemistry and atomic physics,
the Coulomb interaction is of course well-known so there is no intrinsic
need to formulate an effective interaction. However, in lieu of the in
general low order of convergence implied by
Eqn.~(\ref{eq:eigenvalue-estimate}), we believe that HO-based
calculations like the CI method in general may benefit
from the use of effective interactions.

A complete account of the effective interaction scheme outlined here
is out of scope for the present article, but we refer to
Refs.~\onlinecite{Kvaal2007,Kvaal2008a,Kvaal2008b,Navratil2000,Klein1974}
for details as well as numerical algorithms.

Recall, that the interaction is given by
\begin{equation} U = \sum_{i<j}^N C(i,j) = \sum_{i<j}^N
\frac{\lambda}{\|\vec{r}_i-\vec{r}_j\|}, \end{equation} 
a sum of fundamental two-body interactions. For the $N=2$ problem we
have in principle the exact solution, since the
Hamiltonian (\ref{eq:two-electron-ham})
can be reduced to a one-dimensional radial equation, e.g., the
eigenproblem of $H_r$ defined in Eqn.~(\ref{eq:H_r}). This equation may
be solved to arbitrarily high precision using various methods, for
example using a basis expansion in generalized half-range Hermite
functions.\cite{Ball2002}
In nuclear physics, a common approach is to take the best two-body CI
calculations available, where
$R = O(10^3)$, as ``exact'' for this purpose. 

We now define the effective Hamiltonian for $N=2$ as a Hermitian
operator $H_\text{eff}$ defined only within $\Proj_R(N=2)$ that
gives $K = \dim[\Proj_R(N=2)]$ \emph{exact} eigenvalues $E_k$ of $H$, and
$K$ approximate eigenvectors $\psi_{\text{eff},k}$. Of course,
there are infinitely many choices for the $K$ eigenpairs, but by
treating $U = \lambda/r_{12}$ as a perturbation, and ``following'' the
unperturbed HO eigenpairs ($\lambda=0$) through increasing values
of $\lambda$, one makes the eigenvalues unique.\cite{Schucan1973,Kvaal2007} 
The approximate eigenvectors $\psi_{\text{eff},k}\in\Proj_R(N=2)$ are chosen by
minimizing the distance to the exact eigenvectors $\psi_{k}\in
\Hilb(N=2)$ while retaining orthonormality.\cite{Kvaal2008a} This
uniquely defines $H_\text{eff}$ for the two-body system. In terms of
matrices, we have
\begin{equation} H_\text{eff} = \tilde{U}
  \operatorname{diag}(E_1,\cdots,E_K)
  \tilde{U}^\dag, \label{eq:heff-svd} \end{equation}
where $X$ and $Y$ are unitary matrices defined as follows. Let $U$ be the $K\times K$
matrix whose $k$'th column is the coefficients of $P_R\psi_k$. Then
the singular value decomposition of $U$ can be written
\begin{equation} U = X\Sigma Y^\dag, \end{equation}
where $\Sigma$ is diagonal. Then, 
\begin{equation}\tilde{U} := XY^\dag. \end{equation}
The columns of $\tilde{U}$ are the projections $P_R\psi_k$
``straightened out'' to an orthonormal set. Eqn.~(\ref{eq:heff-svd})
is simply the spectral decomposition of $H_\text{eff}$. Although
different in form than most implementations in the literature (e.g.,
Ref.~\onlinecite{Navratil2000}), it is equivalent.

The effective two-body interaction $C_\text{eff}(i,j)$ is now given by
\begin{equation} C_\text{eff}(1,2) := H_\text{eff} - P_R T P_R, \end{equation}
which is defined only within $\Proj_R(N=2)$. 

The $N$-body effective Hamiltonian is defined by
\begin{equation} 
H_\text{eff} := P_RTP_R + \sum_{i<j}^N C_\text{eff}(i,j),
\label{eq:eff-hamiltonian} \end{equation}
where $P_R$ projects onto $\Proj_R(N)$, and thus $H_\text{eff}$ is
defined \emph{only} within
$\Proj_R(N)$. The diagonalization of $H_\text{eff}(N)$ is 
equivalent to a perturbation
technique where a certain class of diagrams is summed to infinite
order in the full problem.\cite{Klein1974} In implementations,
\eqref{eq:heff-svd} and \eqref{eq:eff-hamiltonian} are treated in
COM coordinates, utilizing block 
diagonality of both $H$ and $H_\text{eff}$, see
Ref.~\onlinecite{Kvaal2008b} for details.

We comment that unlike the bare Coulomb interaction, the effective
two-body interaction $C_\text{eff}$ corresponds to a non-local
potential due to the ``straightening out'' of truncated eigenvectors.

Rigorous mathematical treatment of the convergence properties of the
effective interaction is, to the author's knowledge, not
available. Effective interactions have, however, enjoyed great success
in the nuclear physics community, and we strongly believe that we soon
will see sufficient proof of the improved accuracy with this
method. Indeed, in Sec.~\ref{sec:numeric} we see clear evidence of the
accuracy boost when using an effective interaction.

\section{Numerical results}
\setcounter{equation}{0}

\label{sec:numeric}

\subsection{Code description}
\label{sec:code}

We now present numerical results using the full
configuration-interaction method for $N=2$--$5$ electrons in $d=2$
dimensions. We will use both the ``bare'' Hamiltonian $H = T + U$ and
the effective Hamiltonian (\ref{eq:eff-hamiltonian}).

Since the Hamiltonian commutes with angular momentum $L_z$, the latter
taking on eigenvalues $M \in \mathbb{Z}$, the Hamiltonian matrix is
block diagonal. (Recall, that the Fock-Darwin orbitals $\PhiFD_{n,m}$
are eigenstates of $L_z$ with eigenvalue $m$, so each Slater
determinant has eigenvalue $M = \sum_{k=1}^N m_k$.) Moreover, the
calculations are done in a basis of joint eigenfunctions for total
electron spin $S^2$ and its projection $S_z$, as opposed to the Slater
determinant basis used for convergence analysis. Such basis functions
are simply linear combinations of Slater determinants within the same
shell, and further reduce the dimensionality of the Hamiltonian
matrix.\cite{Rontani2006} The eigenfunctions of $H$ are thus
labeled with the total spin $S = 0, 1, \cdots, \frac{N}{2}$ for even
$N$ and $S=\frac{1}{2},\frac{3}{2},\cdots,\frac{N}{2}$ for odd $N$, as
well as the total angular momentum $M = 0, 1, \cdots$. ($-M$ produce
the same eigenvalues as $M$, by symmetry.) We thus split $\Proj_R(N)$ (or
$\Model_R(N)$) into invariant subspaces $\Proj_R(N,M,S)$
($\Model_R(N,M,S)$) and perform computations solely within these.

The calculations were carried out with a code similar to that
described by Rontani \emph{et al.} in
Ref.~\onlinecite{Rontani2006}. Table \ref{tab:code-comparison} shows
comparisons of the present code with that of Table IV of
Ref.~\onlinecite{Rontani2006} for various parameters using the model
space $\Model_R(N,M,S)$. Table \ref{tab:code-comparison} also shows the
case $\lambda=1$, $N=2$, $M=0$, and $S=0$, whose exact lowest
eigenvalue is $E_0 = 3$, cf.~Sec.~\ref{sec:bad-example}. We note that
there are some discrepancies between the results in the last digits of
the results of Ref.~\onlinecite{Rontani2006}. The spaces $\Model_R(N)$
were identical in the two approaches, i.e., the number of basis
functions and the number of non-zero matrix elements produced are
cross-checked and identical.

We have checked that the code also reproduces the results of
Refs.~\onlinecite{Mikhailov2002,Mikhailov2002b,Wensauer2004}, using
the $\Proj_R(N,M,S)$ spaces. Our code is described
in detail elsewhere\cite{Kvaal2008b} where it is also 
demonstrated that it reproduces the eigenvalues of an analytically
solvable $N$-particle system\cite{Johnson1991} to machine precision.

\begin{table*}
\caption{Comparison of current code and
  Ref.~\onlinecite{Rontani2006}\label{tab:code-comparison}. Figures
  from the latter have varying number of significant digits. We include more
  digits from our own computation for reference}
\begin{ruledtabular}
\begin{tabular}{c|ccc|dd|dd|dd}
 &  &  &  & \multicolumn{2}{c}{$R=5$} &
  \multicolumn{2}{c}{$R=6$} & \multicolumn{2}{c}{$R=7$} \\
$N$ & $\lambda$ & $M$ & $2S$ &\multicolumn{1}{c}{Current} & \multicolumn{1}{c}{Ref.~\onlinecite{Rontani2006}}
&\multicolumn{1}{c}{Current}  & \multicolumn{1}{c}{Ref.~\onlinecite{Rontani2006}}
&\multicolumn{1}{c}{Current}  & \multicolumn{1}{c}{Ref.~\onlinecite{Rontani2006}} \\
\hline
2 & 1 & 0 & 0 & 3.013626 &        & 3.011020 &        & 3.009236 &      \\
  & 2 & 0 & 0 & 3.733598 & 3.7338 & 3.731057 & 3.7312 & 3.729324 &3.7295\\
  &   & 1 & 2 & 4.143592 & 4.1437 & 4.142946 & 4.1431 & 4.142581 &4.1427\\
3 & 2 & 1 & 1 & 8.175035 & 8.1755 & 8.169913 &        & 8.166708 &8.1671\\
  & 4 & 1 & 1 & 11.04480 & 11.046 & 11.04338 &        & 11.04254 &11.043\\
  &   & 0 & 3 & 11.05428 & 11.055 & 11.05325 &        & 11.05262 &11.053\\
4 & 6 & 0 & 0 & 23.68944 & 23.691 & 23.65559 &        & 23.64832 &23.650\\
  &   & 2 & 4 & 23.86769 & 23.870 & 23.80796 &        & 23.80373 &23.805\\
5  & 2  & 0 & 5 & 21.15093 & 21.15  & 21.13414 & 21.13  & 21.12992        &21.13 \\
  & 4  & 0 & 5 & 29.43528 & 29.44  & 29.30898 & 29.31  & 29.30251 &29.30\\
\end{tabular}   
\end{ruledtabular}
\end{table*}

\subsection{Experiments}
\label{sec:experiments}

\begin{figure}
\includegraphics[width=\columnwidth]{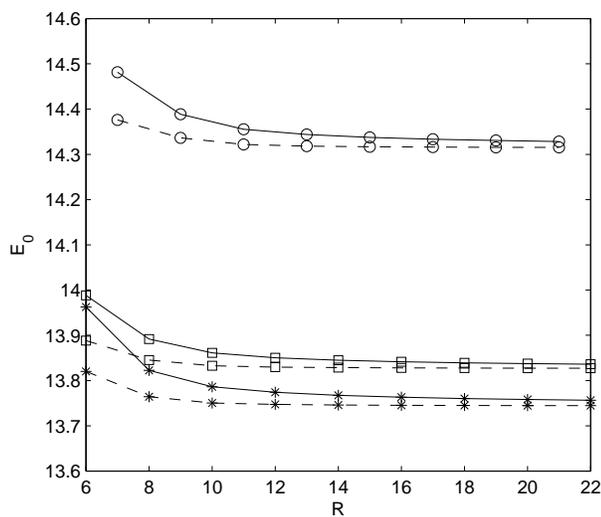}%
\caption{\label{fig:raw-plots}Eigenvalues for $N=4$, $S=0$, $\lambda =
  2$ as function of $R$ for $H$ (solid) and $H_\text{eff}$
  (dashed). $M=0,1$ and $2$ are represented by squares, circles and
  stars, resp.}
\end{figure}

For the remainder, we only use the energy cut spaces $\Proj_R(N,M,S)$.
Figure \ref{fig:raw-plots} shows the development of the lowest
eigenvalue $E_0 = E_0(N,M,S)$ for $N=4$, $M=0,1,2$ and $S=0$ as
function of the shell truncation parameter $R$, using both
Hamiltonians $H$ and $H_\text{eff}$. Apparently, the effective
interaction eigenvalues provide estimates for the ground state
eigenvalues that are better than the bare interaction
eigenvalues. This effect is attenuated with higher $N$, due to the
fact that the two-electron effective Coulomb interaction does not take
into account three- and many-body effects which become substantial for
higher $N$.

\begin{figure*}
\subfigure{\includegraphics[width=0.45\textwidth]{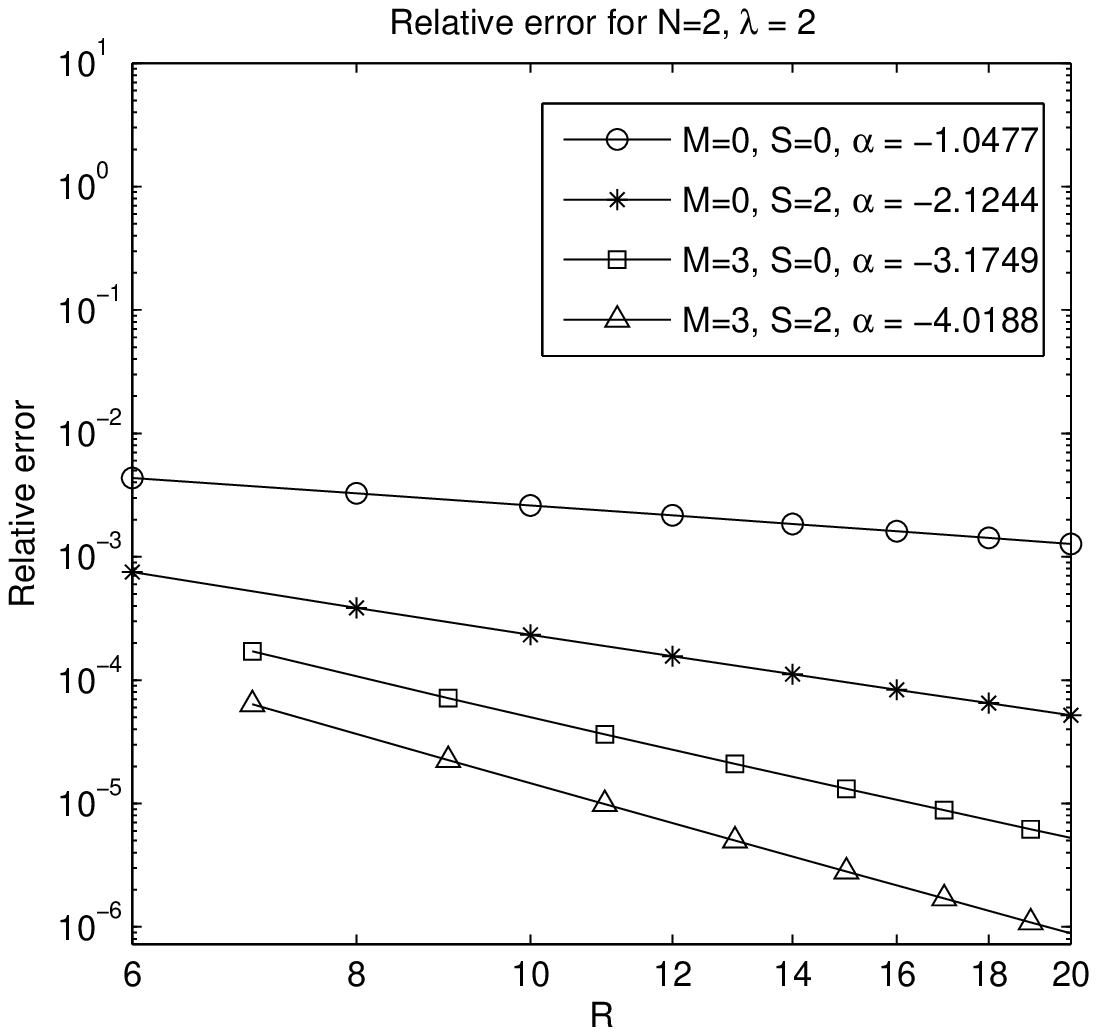}}
\subfigure{\includegraphics[width=0.45\textwidth]{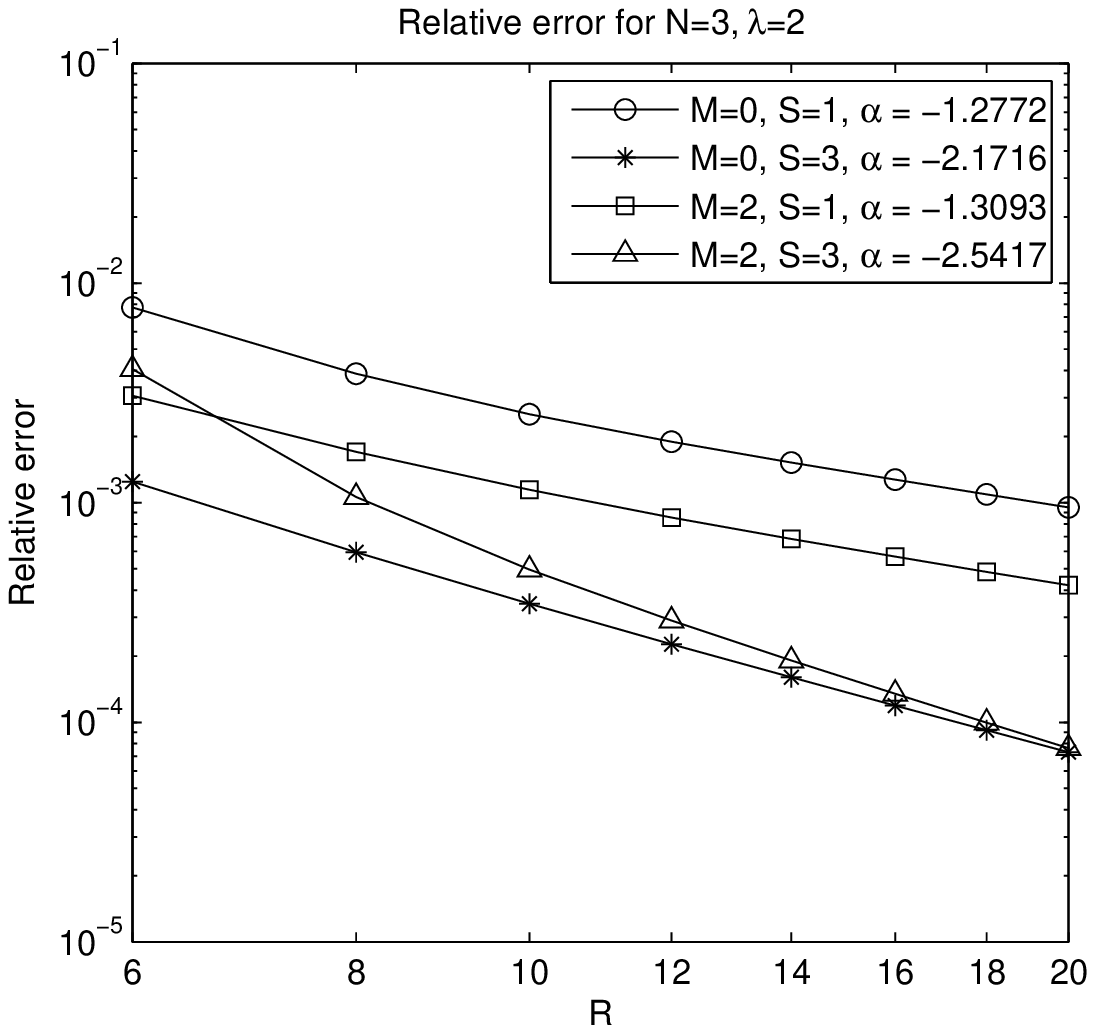}}
\subfigure{\includegraphics[width=0.45\textwidth]{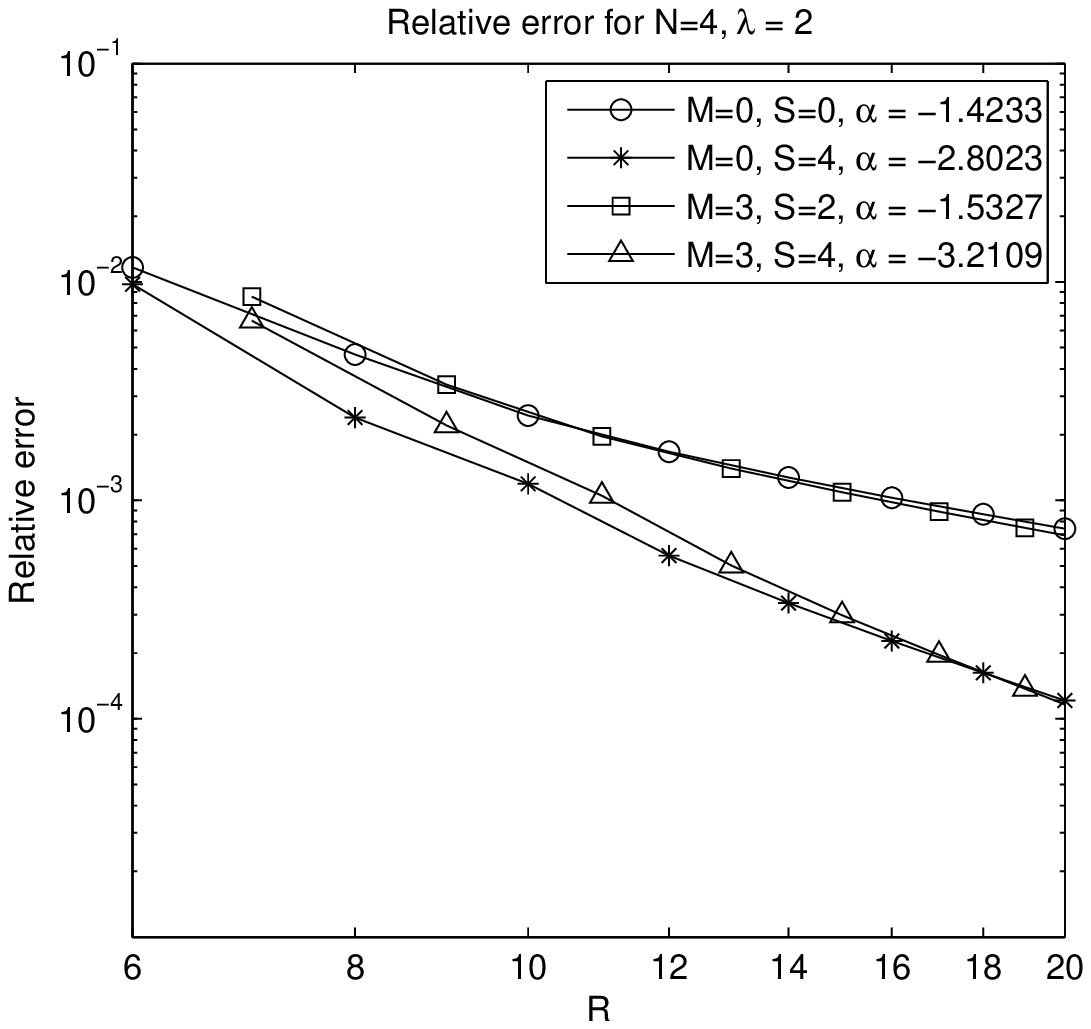}}
\subfigure{\includegraphics[width=0.45\textwidth]{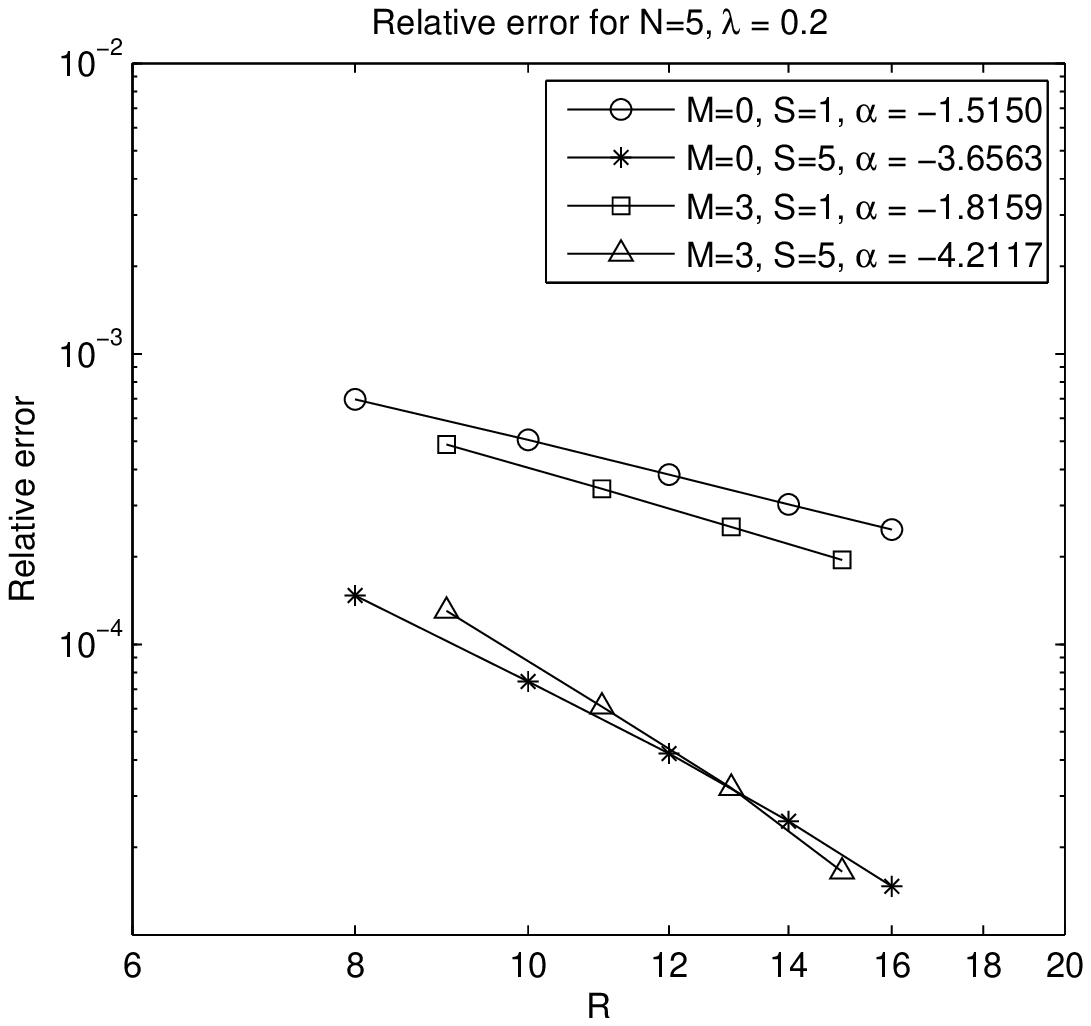}}
\caption{\label{fig:loglog-error}Plot of relative error using the bare
  interaction for various
  $N$, $M$ and $S$. Clear $o(R^{\alpha})$ dependence in all cases.}
\end{figure*}

We take the $H_\text{eff}$-eigenvalues as
``exact'' and graph the relative error in $E_0(N,M,S)$ as
function of $R$ on a logarithmic scale in Fig.~\ref{fig:loglog-error}, in anticipation of the
relation
\begin{equation} \ln(E_h-E) \approx C + \alpha\ln R, \quad \alpha =
  -(k-1+\epsilon). \end{equation}
The graphs show straight lines for large
$R$, while for small $R$ there is a transient region of non-straight
lines. For $N=5$, however, $\lambda = 2$ is too large a value to
reach the linear regime for the range of $R$ available, so in this
case we chose to plot the corresponding error for the very small value
$\lambda = 0.2$, showing clear straight lines in the error. The slopes
are more or less independent of $\lambda$, as observed in different
calculations. 

\begin{figure*}
\subfigure{\includegraphics[width=0.45\textwidth]{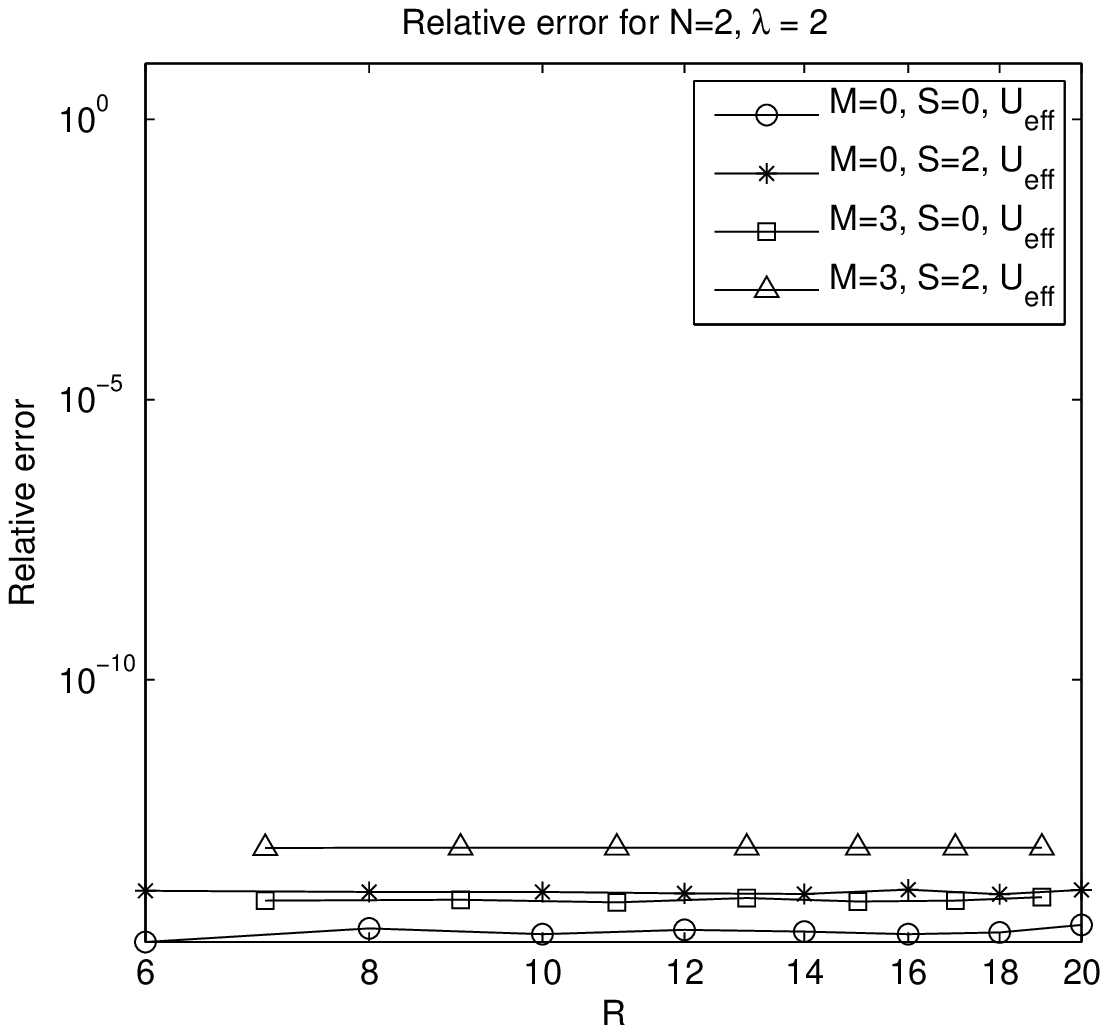}}
\subfigure{\includegraphics[width=0.45\textwidth]{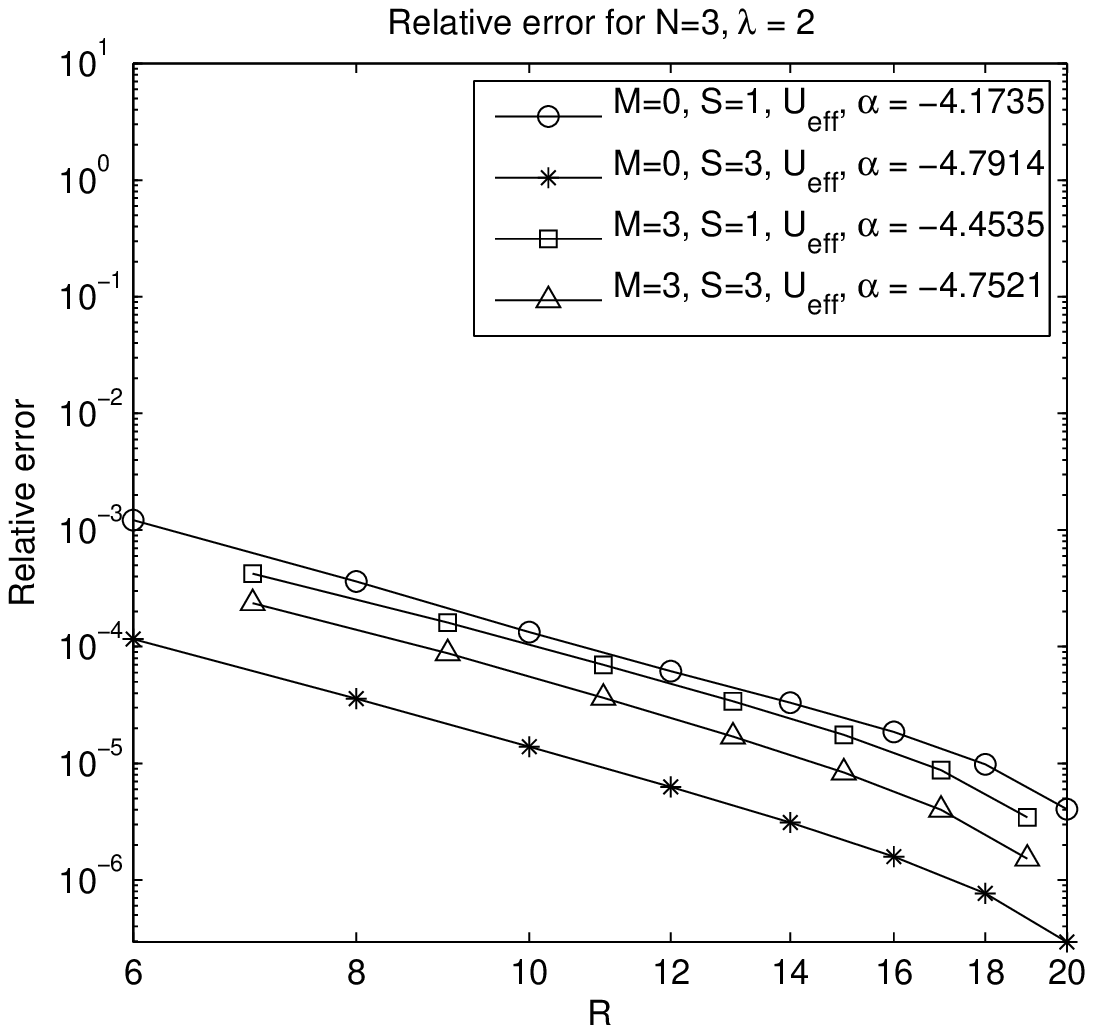}}
\subfigure{\includegraphics[width=0.45\textwidth]{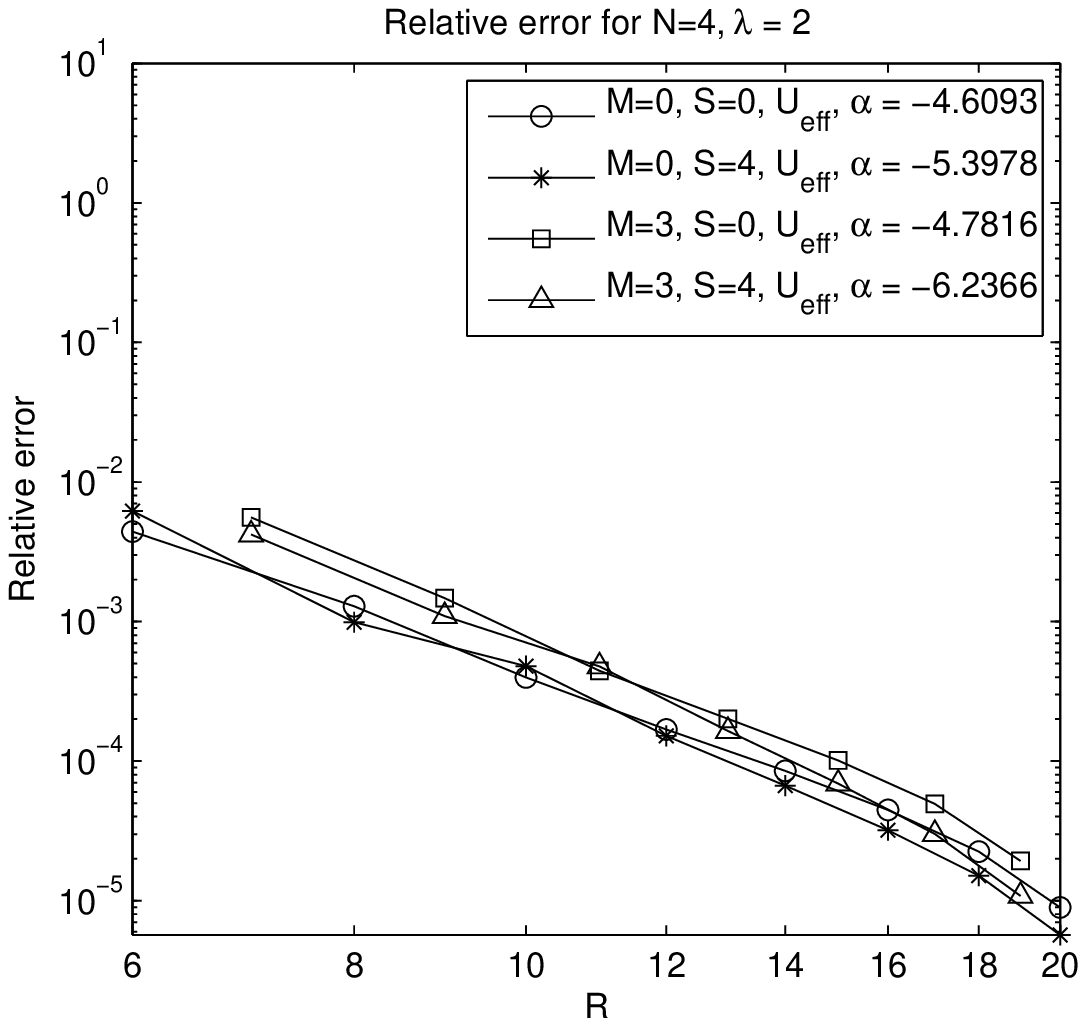}}
\subfigure{\includegraphics[width=0.45\textwidth]{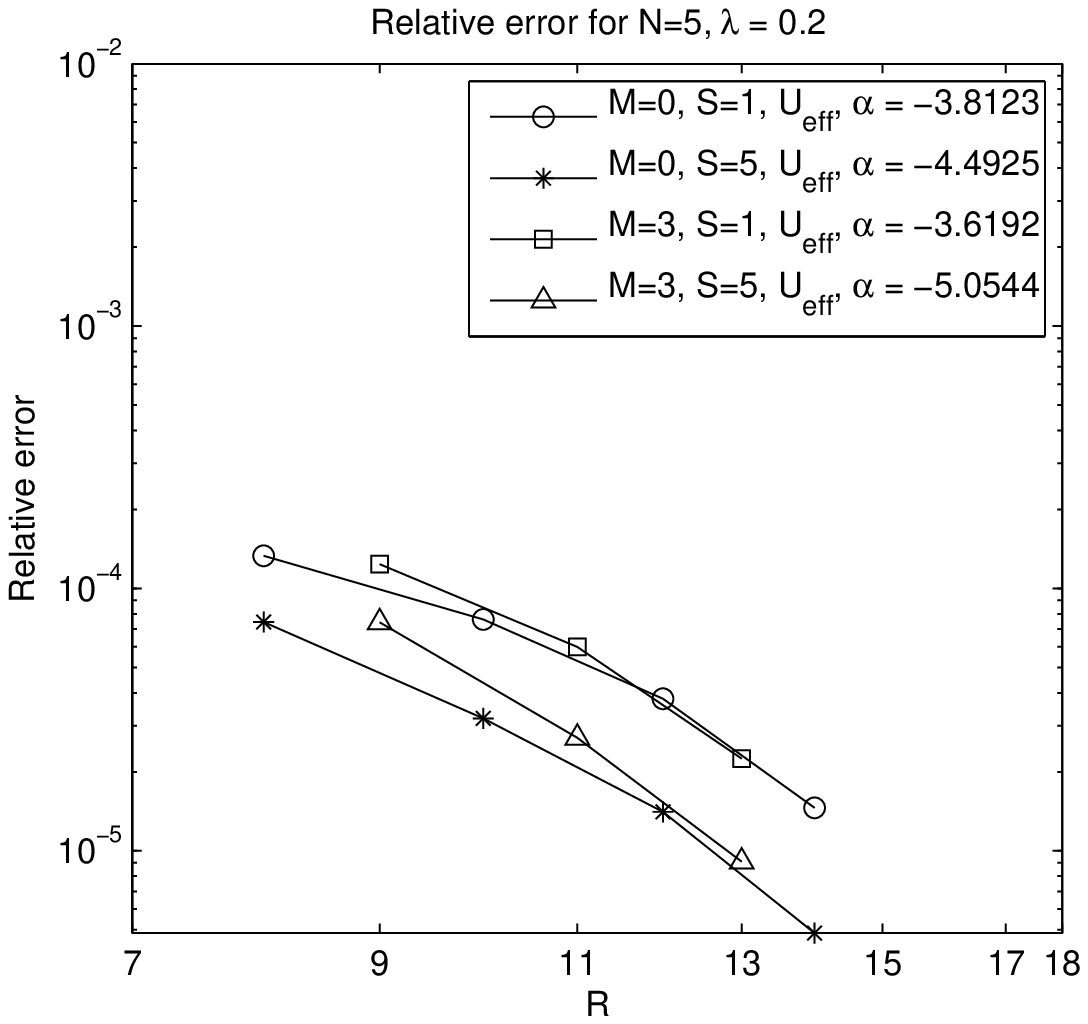}}
\caption{\label{fig:loglog-error-eff}Plot of relative error using an
  effective interaction for various
  $N$, $M$ and $S$. Clear $o(R^{\alpha})$ dependence in all cases, but
notice artifacts when $R$ is large, due to errors in most correct
eigenvalues. The $R=5$ case does not contain enough data to compute
the slopes with sufficient accuracy.}
\end{figure*}

In Fig.~\ref{fig:loglog-error-eff} we show the corresponding graphs
when using the effective Hamiltonian $H_\text{eff}$. We estimate the
relative error as before, leading to artifacts for the largest values
of $R$ due to the fact that there is a finite error in the best estimates
for the eigenvalues. However, in all cases there are clear, linear
regions, in which we estimate the slope $\alpha$. In all cases, the
slope can be seen to decrease by at least $\Delta\alpha\approx -1$ compared to
Fig.~\ref{fig:loglog-error}, indicating that the effective interaction
indeed accelerates the CI convergence by at least an order of
magnitude. We also observe, that the relative errors are improved by
an order of magnitude or more for the lowest values of $R$ shown,
indicating the gain in accuracy when using small model spaces with the
effective interaction.

Notice, that for symmetry reasons only even (odd) $R$ for even (odd)
$M$ yields increases in basis size $\dim[\Proj_R(N,M,S)]$, so only
these values are included in the plots.

To overcome the limitations of the two-body effective interaction for
higher $N$, an effective three-body interaction could be considered,
and is hotly debated in the nuclear physics community. (In nuclear
physics, there are also more complicated three-body effective forces
that need to be included.\cite{Caurier2005}) However, this will lead
to a huge increase in memory consumption due to extra nonzero matrix
elements. At the moment, there are no methods available that can
generate the exact three-body effective interaction with sufficient
precision.

We stress, that the relative error decreases very slowly in
general. It is a common misconception, that if a number of digits of
$E_0(N,M,S)$ is unchanged between $R$ and $R+2$, then these digits
have converged. This is not the case, as is easily seen from
Fig.~\ref{fig:loglog-error}. Take for instance $N=4$, $M=0$ and $S=0$,
and $\lambda = 2$. For $R=14$ and $R=16$ we have $E_0 = 13.84491$ and
$E_0 = 13.84153$, respectively, which would give a relative error
estimate of $2.4\times 10^{-4}$, while the correct relative error is
$1.3\times 10^{-3}$.

The slopes in Fig.~\ref{fig:loglog-error} vary greatly,
showing that the eigenfunctions indeed have varying global smoothness,
as predicted in Sec.~\ref{sec:analytic-props}. For
$(N,M,S)=(5,3,5/2)$, for example, $\alpha \approx -4.2$, indicating
that $\psi\in H^5(\mathbb{R}^{10})$. It seems, that higher $S$ gives
higher $k$, as a rule of thumb. Intuitively, this is because the Pauli
principle forces the wave function to be zero at coalesce points,
thereby generating smoothness.

\section{Discussion and conclusion}
\setcounter{equation}{0}

\label{sec:conclusion}

We have studied approximation properties of Hermite functions and
harmonic oscillator eigenfunctions. This in turn allowed for a
detailed convergence analysis of numerical methods such as the CI
method for the parabolic quantum dot. Our main conclusion, is, that for wave
functions $\psi\in H^k(\mathbb{R}^n)$ falling off exponentially as
$\|x\|\rightarrow\infty$, the shell-weight function $p(r)$ decays as
$p(r) = o(r^{-k-1})$. Applying this to the convergence theory of the
Ritz-Galerkin method, we obtained the estimate
(\ref{eq:CI-eigenvalue-error}) for the error in the eigenvalues. 
A complete characterization of the upper bound on the differentiability
$k$, i.e., in $\psi^{(s)}\in H^k$, as well as a study of the constant
$K$ in Eqn.~(\ref{eq:eigenvalue-estimate}), would complete our
knowledge of the convergence of the CI calculations.

We also demonstrated numerically, that the use of a two-body effective
interaction accelerates the convergence by at least an order of
magnitude, which shows that such a method \emph{should} be used
whenever possible. On the other hand, a rigorous mathematical study of
the method is yet to come. Moreover, we have not investigated to what
extent the increase in convergence is independent of the interaction
strength $\lambda$. This, together with a study of the accuracy of the
\emph{eigenvectors}, is an obvious candidate for further
investigation.

The theory and ideas presented in this article should in principle be
universally applicable. In fact, Figs.~1--3 of
Ref.~\onlinecite{Navratil2000} clearly indicates this, where the eigenvalues
of $^3\mathrm{He}$ as function of model space size are graphed both for bare
and effective interactions, showing some of the features we have
discussed.

Other interesting future studies would be a direct comparison of the direct
product model space $\Model_R(N)$ and our energy cut model space
$\Proj_R(N)$. Both techniques are common, but may have different
numerical characteristics. Indeed, $\dim[\Model_R(N)]$ grows much
quicker than $\dim[\Proj_R(N)]$, while we are uncertain of whether the
increased basis size yields a corresponding increased
accuracy. 

We have focused on the parabolic quantum dot, firstly because it
requires relatively small matrices to be stored, due to conservation
of angular momentum, but also because it is a widely studied
model. Our analysis is, however, general, and applicable to other
systems as well, e.g., quantum dots trapped in double-wells, finite
wells, and so on. Indeed, by adding a one-body potential $V$ to the
Hamiltonian $H=T+U$ we may model other geometries, as well as adding
external fields.\cite{Helle2005,Brasken2000,Corni2003}

\section*{Acknowledgments}

The author wishes to thank Prof.~M.~Hjorth-Jensen (CMA) for useful
discussions, suggestions and feedback, and also Dr.~G.~M.~Coclite
(University of Bari, Italy) for mathematical discussions. This work was financed by CMA through
the Norwegian Research Council.

\appendix

\section{Mathematical details}
\setcounter{equation}{0}
\renewcommand{\theequation}{\thesection.\arabic{equation}}

\subsection{Multi-indices}
\label{sec:multi-indices}

A very handy tool for compact and unified notation when the dimension $n$
of the underlying measure space $\mathbb{R}^n$ is a parameter, are
multi-indices. The set $\Ind_n$ of multi-indices are defined as
$n$-tuples of non-negative indices, viz, $\alpha =
(\alpha_1,\cdots,\alpha_n)$, where $\alpha_k\geq 0$.

We define several useful operations on multi-indices as follows. Let
$u$ be a formal vector of $n$ symbols. Moreover, let
$\phi(\xi)=\phi(\xi_1,\xi_2,\cdots,\xi_n)$ be a function. Then, define
\begin{eqnarray}
|\alpha| &\equiv& \alpha_1 + \alpha_2 + \cdots + \alpha_n \\
\alpha! &\equiv&\alpha_1!\alpha_2!\cdots\alpha_n! \\
\alpha \pm \beta &\equiv&(\alpha_1 \pm \beta_1,\cdots,\alpha_n \pm
\beta_n) \label{eq:multi-index-add} \\
u^\alpha &\equiv& u_1^{\alpha_1}u_2^{\alpha_2}\cdots
u_n^{\alpha_n}, \label{eq:multi-index-power} \\
\partial^\alpha\phi(\xi) &\equiv& \pdiff{^{\alpha_1}}{\xi_1^{\alpha_1}}
\pdiff{^{\alpha_1}}{\xi_2^{\alpha_2}} \cdots
\pdiff{^{\alpha_d}}{\xi_d^{\alpha_1}}\phi(\xi)  \label{eq:multi-index-diff}
\end{eqnarray}
In Eqn.~\eqref{eq:multi-index-add}, the result may not be a
multi-index when we subtract two indices, but this will not be an issue for us. Notice, that
Eqn.~(\ref{eq:multi-index-diff}) is a mixed partial derivative of order $|\alpha|$.
Moreover, we say that $\alpha < \beta$ if and only if
$\alpha_j<\beta_j$ for all $j$. We define $\alpha=\beta$ similarly. We also define ``basis
indices'' $e_j$ by $(e_j)_{j'} = \delta_{j,j'}$.  We comment,
that we will often use the notation $\partial_x \equiv \frac{\partial}{\partial
  x}$ and $\partial_k \equiv \frac{\partial}{\partial
  {\xi_k}}$ to simplify notation. Thus,
\begin{equation}
  \partial^\alpha = (\partial^{\alpha_1}_1, \cdots,
\partial^{\alpha_n}_n),
\end{equation}
consistent with Eqn.~(\ref{eq:multi-index-power}).

\subsection{Weak derivatives and Sobolev spaces}
\label{sec:sobolev}

We present a quick summary of weak derivatives and related concepts
needed. The material is elementary and superficial, but probably
unfamiliar to many readers, so we include it here. Many terms will be
left undefined; if needed, the reader may consult standard texts,
e.g., Ref.~\onlinecite{Evans1998}.

The space $L^2(\mathbb{R}^n)$ is defined as
\begin{equation}\label{eq:L2} L^2(\mathbb{R}^n) \equiv \left\{ \psi :
  \mathbb{R}^n\rightarrow \mathbb{C} \; : \; \int_{\mathbb{R}^n}
  |\psi(\xi)|^2 \rmd^n\xi \; < +\infty \right\}, \end{equation}
where the Lebesgue integral is more general than the Riemann
``limit-of-small-boxes'' integral. It is important that we identify
two functions $\psi$ and $\psi_1$ differing only at a set
$Z\in\mathbb{R}^n$ of measure zero. Examples of such sets are points
if $n\geq 1$, curves if $n\geq 2$, and so on, and countable unions of
such. For example, the rationals constitute a set of measure zero in
$\mathbb{R}$. Under this assumption, $L^2(\mathbb{R}^n)$ becomes a
Hilbert space with the inner product
\begin{equation} 
  \braket{\psi_1, \psi_2} \equiv \int_{\mathbb{R}^n} \psi_1(\xi)^\ast
  \psi_2(\xi) \rmd^n\xi, 
\end{equation}
where the asterisk denotes complex conjugation.

The classical derivative is too limited a concept for the abstract
theory of partial differential equations, including the Schr\"odinger
equation. Let $\mathcal{C}^\infty_0$ be the set of infinitely
differentiable functions which are non-zero only in a ball of finite
radius. Of course, $\mathcal{C}^\infty_0\subset
L^2(\mathbb{R}^n)$. Let $\psi\in L^2(\mathbb{R}^n)$, and let
$\alpha\in \Ind_n$ be a multi-index. If there exists a $v\in
L^2(\mathbb{R}^n)$ such that, for all $\phi\in \mathcal{C}^\infty_0$,
\begin{equation}
  \int_{\mathbb{R}^n} (\partial^\alpha \phi(\xi)) \psi(\xi) \rmd^n \xi =
  (-1)^{|\alpha|} \int_{\mathbb{R}^n} \phi(\xi) v(\xi) \rmd^n \xi, 
\end{equation}
then $\partial^\alpha\psi \equiv v \in L^2$ is said to be a \emph{weak
  derivative}, or distributional derivative, of $\psi$. In this way,
the weak derivative is defined in an average sense, using integration
by parts.

The weak derivative is unique (up to redefinition on a set of measure
zero), obeys the product rule, chain rule, etc.

It is easily seen, that if $\psi$ has a classical derivative $v\in
L^2(\mathbb{R}^n)$ it coincides with the weak derivative. Moreover, if
the classical derivative is defined almost everywhere (i.e.,
everywhere except for a set of measure zero), then $\psi$ has a weak
derivative.

The Sobolev space $H^k(\mathbb{R}^n)$ is defined as the subset of
$L^2(\mathbb{R}^n)$ given by
\begin{equation}\label{eq:Hk} H^k(\mathbb{R}^n) \equiv \left\{ \psi\in L^2
  \; : \; \partial^\alpha\psi \in L^2, \quad \forall
  \alpha\in\Ind_n,\;|\alpha|\leq k \right\}.  \end{equation}
The Sobolev space is also a Hilbert space with the inner product
\begin{equation} (\psi_1,\psi_2) \equiv \sum_{\alpha,\;|\alpha|\leq k}
\braket{\partial^\alpha\psi_1,\partial^\alpha\psi_2}, \end{equation}
and this is the main reason why one obtains a unified theory of
partial differential equations using such spaces.

The space $H^k(\mathbb{R}^n)$ for $n>1$ is a big space -- there are
some exceptionally ill-behaved functions there, for example there are
functions in $H^k$ that are unbounded on arbitrary small regions but
still differentiable. (Hermite series for such functions would still
converge faster than, e.g., for a function with a jump discontinuity!)
For our purposes, it is enough to realize that the Sobolev spaces
offer exactly the notion of derivative we need in our analysis of the
Hermite function expansions. 

\subsection{Proofs of propositions}
\label{app:proofs}
\setcounter{theorem}{0}

We will now prove the propositions given in Sec.~\ref{sec:series-new},
and also discuss these results on mathematical terms. 

Recall that the Hermite functions $\phi_n\in L^2(\mathbb{R})$, where
$n\in\mathbb{N}_0$ is a non-negative integer, are defined by
\begin{equation} 
  \phi_n(x) = (2^n n! \sqrt{\pi})^{-1/2} H_n(x)
  e^{-x^2/2}, \label{eq:hermitefunc} 
\end{equation} 
where $H_n(x)$ is the usual Hermite polynomial.

A well-known method for finding the eigenfunctions of $H_{\text{HO}}$
in one dimension involves writing
\begin{equation} H_{\text{HO}} = a^\dag a + \frac{1}{2}, \end{equation}
where the ladder operator $a$ is given by
\begin{equation} a \equiv \frac{1}{\sqrt{2}}(x + \partial_x),  \end{equation}
with Hermitian adjoint $a^\dag$ given by
\begin{equation} a^\dag \equiv \frac{1}{\sqrt{2}}(x - \partial_x). \end{equation}
The name ``ladder operator'' comes from the important formulas
\begin{eqnarray} 
a \phi_n(x) &=& \sqrt{n}\phi_{n-1}(x) \label{eq:ladder}\\
a^\dag \phi_n(x) &=& \sqrt{n+1}\phi_{n+1}(x), \label{eq:ladder2}
\end{eqnarray}
valid for all $n$. This can easily be proved by using the recurrence
relation \eqref{eq:recurrence}. By repeatedly acting on $\phi_0$ with
$a^\dag$ we generate every Hermite function, viz,
\begin{equation} \phi_n(x) = n!^{-1/2} (a^\dag)^n
  \phi_0(x). \label{eq:generate-hermite} \end{equation}

As Hermite functions constitute a complete, orthonormal sequence in
$L^2(\mathbb{R})$, any $\psi\in L^2(\mathbb{R})$ can be written
as a series in Hermite functions, viz,
\begin{equation} \psi(x) = \sum_{n=0}^\infty c_n \phi_n(x), \end{equation}
where the coefficients $c_n$ are uniquely determined by $c_n = \langle
\phi_n, \psi \rangle$. 

An interesting fact is that the Hermite functions are
also eigenfunctions of the Fourier transform with eigenvalues
$(-i)^n$, as can easily be proved by induction by observing firstly that
the Fourier transform of $\phi_0(x)$ is $\phi_0(k)$ itself, and secondly
that the Fourier transform of $a^\dag$ is $-i a^\dag$ (acting on the
variable $y$). It follows from
completeness of the Hermite functions, that the Fourier transform defines a
unitary operator on $L^2(\mathbb{R})$.

We now make a simple observation, namely that
\begin{equation} (a^\dag)^k \phi_n(x) = P_k(n)^{1/2}
  \phi_{n+k}(x), \label{eq:basic-fact} \end{equation}
where $P_k(n)=(n+k)!/n!>0$ is a polynomial of degree $k$
for $n\geq 0$. Moreover $P_k(n+1) > P_k(n)$ and $P_{k+1}(n) > P_k(n)$.

We now prove the following lemma:
\begin{lemma}[Hermite series in one dimension]\label{lemma:approx1}
Let $\psi\in L^2(\mathbb{R})$. Then
\begin{enumerate}
\item $a^\dag\psi\in L^2(\mathbb{R})$ if and only if $a\psi\in
  L^2(\mathbb{R})$  if and only if $\sum_{n=0}^\infty n|c_n|^2 <
  +\infty$, where $c_n = \langle \phi_n, \psi \rangle$ \label{point1}
\item $a^\dag\psi\in L^2(\mathbb{R})$ if and only if
  $x\psi,\partial_x\psi\in L^2(\mathbb{R})$ \label{point1b}
\item $(a^\dag)^{k+1}\psi\in L^2(\mathbb{R})$ implies $(a^\dag)^k\psi\in L^2(\mathbb{R})$ \label{point2}
\item $(a^\dag)^k\psi\in L^2(\mathbb{R})$ if and only if
\begin{equation} \sum_{n=0}^\infty n^k|c_n|^2 <
  +\infty. \label{eq:weighted-sum} \end{equation} \label{point3}
\item $(a^\dag)^k\psi\in L^2(\mathbb{R})$ if and only if
  $x^j\partial_x^{k-j}\psi\in L^2(\mathbb{R})$ for $0\leq j \leq k$  \label{point4}
\end{enumerate}
\end{lemma}

\emph{Proof:} We have
\begin{equation} \|a^\dag \psi\|^2 = \sum_{n=0}^\infty (n+1)|c_n|^2 = \|\psi^2\| +
\|a\psi\|^2, \end{equation} from which statement \ref{point1} follows. Statement
\ref{point1b} follows from the definition of $a^\dag$, and that
$a^\dag\psi\in L^2$ implies $a\psi\in L^2$ (since $\|a\psi\|\leq\|a^\dag\psi\|$), which again implies
$x\psi,\partial_x\psi\in L^2$. Statement \ref{point2} follows from the
monotone behaviour of $P_k(n)$ as function of $k$. Statement
\ref{point3} then follows. By iterating statement \ref{point1b} and
using $[\partial_x,x]=1$ and statement \ref{point2}, statement
\ref{point4} follows.\eofproof

The significance of the condition $a^\dag \psi\in L^2(\mathbb{R})$ is
that the coefficients $c_n$ of $\psi$ must decay faster than for a
completely arbitrary wave function in $L^2(\mathbb{R})$. Moreover,
$a^\dag\psi\in L^2(\mathbb{R})$ is the same as requiring
$\partial_x\psi\in L^2(\mathbb{R})$, and $x\psi\in
L^2(\mathbb{R})$. Lemma \ref{lemma:approx1} also generalizes this
fact for $(a^\dag)^k\psi\in L^2(\mathbb{R})$, giving successfully
quicker decay of the coefficients. In all cases, the decay is 
expressed in an average sense, through a growing weight function in a
sum, as in Eqn.~(\ref{eq:weighted-sum}). Since the terms in the sum
must converge to zero, this implies a pointwise faster decay, as
stated in Eqn.~(\ref{eq:order-1d}) below.

We comment here, that the partial derivatives must be understood in
the weak, or distributional, sense: Even though $\psi$ may not be
everywhere differentiable in the ordinary sense, it may have a weak
derivative.  For example, if the classical derivative exists
everywhere except for a countable set (and if it is in $L^2$), this is
the weak derivative.  Moreover, if this derivative has a jump
discontinuity, there are no higher order weak derivatives. 

Loosely speaking, since $x\psi(x)\in L^2\Leftrightarrow
\partial_y\hat{\psi}(y)\in L^2$, where $\hat{\psi}(y)$ is the Fourier
transform, point \ref{point1b} of Lemma \ref{lemma:approx1} is a
combined smoothness condition on $\psi(x)$ and $\hat{\psi}(y)$. Point
\ref{point4} is a generalization to higher derivatives, but is
difficult to check in general for an arbitrary $\psi$. On the other
hand, it is well-known that the eigenfunctions of many Hamiltonians of
interest, such as the quantum dot Hamiltonian \eqref{eq:hamiltonian},
decay exponentially fast as $|x|\rightarrow\infty$. For such
exponentially decaying functions over $\mathbb{R}^1$, $x^k\psi\in
L^2(\mathbb{R})$ for all $k\geq 0$, i.e., $\hat{\psi}(y)$ is infinitely
differentiable. We then have the following lemma:
\begin{lemma}[Exponential decay in 1D]\label{lemma:exp1}
Assume that $x^k\psi\in L^2(\mathbb{R})$ for all $k\geq 0$. Then a
sufficient criterion for $(a^\dag)^m\psi\in L^2(\mathbb{R})$ is
$\partial_x^m\psi\in L^2(\mathbb{R})$, i.e., $\psi\in H^m(\mathbb{R})$. In fact,
$x^k\partial_x^{m'}\psi\in L^2$ for all $m'<m$ and all $k\geq 0$.
\end{lemma}

\emph{Proof:} We prove the Proposition inductively. We note that,
since $\partial_x^m\psi\in L^2$ implies $\partial_x^{m-1}\psi\in L^2$, the
proposition holds for $m-1$ if it holds for a given $m$. Moreover, it
holds trivially for $m=1$.

Assume then, that it holds for a given $m$, i.e., that
$\psi\in H^m$ implies $x^k\partial_x^{m-j}\psi\in L^2$ for $1\leq j \leq m$ and for all
$k$ (so that, in particular, $(a^\dag)^m\psi\in L^2$). It remains to
prove, that $\psi\in H^{m+1}$ implies $x^k\partial_x^{m}\psi\in L^2$, since
then $(a^\dag)^{m+1}\psi\in L^2$ by statement \ref{point4} of Lemma
\ref{lemma:approx1}. We compute the norm and use integration by parts,
viz,
\begin{eqnarray*} \|x^k\partial_x^m\psi\|^2 &=& \int_{\mathbb{R}}
x^{2k}\partial_x^k\psi^\ast(x)\partial_x^k\psi(x) \\ &=& - 2k\langle \partial_x^{m}\psi,
x^{2k-1}\partial_x^{m-1}\psi\rangle \\ & & - \langle
\partial_x^{m+1}\psi, x^{2k}\partial_x^{m-1}\psi\rangle < +\infty. 
\end{eqnarray*}
The boundary terms vanish. Therefore,
$x^k\partial_x^m\psi\in L^2$ for all $k$, and the proof is complete.
\eofproof

The proposition states that for the subset of $L^2(\mathbb{R})$ consisting of
exponentially decaying functions, the approximation properties of the
Hermite functions will \emph{only} depend on the smoothness properties of
$\psi$. Moreover, the derivatives up to the penultimate order
decay exponentially as well. (The highest order derivative may
decay much slower.)

From Lemmas \ref{lemma:approx1} and \ref{lemma:exp1} we extract the
following important characterization of the approximating properties
of Hermite functions in $d=1$ dimensions:

\begin{theorem}[Approximation in one dimension]
Let $k \geq 0$ be a given integer. Let $\psi \in L^2(\mathbb{R})$ be
given by
\begin{equation} \psi(x) = \sum_{n=0}^\infty c_n \phi_n(x). \end{equation}
Then $\psi\in H^k(\mathbb{R})$ if and only if
\begin{equation} \sum_{n=0}^\infty n^k |c_n|^2 < \infty. \end{equation}
The latter implies that 
\begin{equation} |c_n| =
  o(n^{-(k+1)/2}). \label{eq:order-1d} \end{equation}
Let $\psi_R = P_R\psi = \sum_{n=0}^R c_n\phi_n$. Then
\begin{equation} \| \psi - \psi_R \| = \left( \sum_{n=R+1}^\infty |c_n|^2
\right)^{1/2}. \end{equation}
\end{theorem}

This is the central result for Hermite series approximation in
$L^2(\mathbb{R}^1)$. Observe that Eqn.~(\ref{eq:order-1d}) implies
that the error $\|\psi-\psi_R\|$ can easily be estimated.
See also Prop.~\ref{prop:approx-general-n} and comments thereafter.

Now a word on pointwise convergence of the Hermite series.
As the Hermite functions are uniformly
bounded,\cite{Boyd1984} viz,
\begin{equation} |\phi_n(x)| \leq 0.816 \quad \forall \; x\in\mathbb{R}, \end{equation}
the pointwise error in $\psi_R$ is bounded by
\begin{equation} |\psi(x) - \psi_R(x)| \leq 0.816 \sum_{n=R+1}^\infty
  |c_n|. \end{equation} 
Hence, if the sum on the right hand side is finite, the convergence is
uniform. If the coefficients $c_n$ decay rapidly enough, both errors
can be estimated by the dominating neglected coefficients.

We now consider expansions of functions in $L^2(\mathbb{R}^n)$. To
stress that $\mathbb{R}^n$ may be other than the configuration space
of a single particle, we use the notation
$x=(x_1,\cdots,x_n)\in\mathbb{R}^n$ instead of
$\vec{r}\in \mathbb{R}^d$. For $N$ electrons in $d$ spatial
dimensions, $n=Nd$.

Recall that the Hermite functions over $\mathbb{R}^n$ are indexed by
multi-indices $\alpha = (\alpha_1,\cdots,\alpha_n)\in \Ind_n$,
and that
\begin{equation} \Phi_{\alpha}(x) \equiv \phi_{\alpha_1}(x_1)\cdots\phi_{\alpha_n}(x_n).
\end{equation}
Now, to each spatial coordinate $x_k$ define the
ladder operators $a_k \equiv (x_k + \partial_k)/\sqrt{2}$. These obey
$[a_j,a_k]=0$ and $[a_j,a^\dag_k]=\delta_{j,k}$, as can easily be verified. Let $\mathbf{a}$ be
a formal vector of the ladder operators, viz,
\begin{equation} \mathbf{a} \equiv (a_1, a_2, \dots, a_n). \end{equation}
For the first Hermite function, we have
\begin{equation} \Phi_0(x) \equiv \pi^{-n/4} e^{-\|x\|^2/2}. \end{equation}
By using
Eqn.~(\ref{eq:generate-hermite}) and
Eqn.~(\ref{eq:multi-index-power}), we may generate all other Hermite
functions, viz,
\begin{equation} \Phi_\alpha(x) \equiv \alpha!^{-1/2}(\mathbf{a}^\dag)^\alpha \Phi_0(x). \label{eq:generate-hermite2}\end{equation}

Given two multi-indices $\alpha$ and $\beta$, we define the polynomial
$P_\alpha(\beta)$ by
\begin{equation} P_\beta(\alpha) \equiv \frac{(\alpha+\beta)!}{\alpha!} = \prod_{j=1}^D
P_{\beta_j}(\alpha_j), \end{equation}
where $P$ is defined for integers as before.

Since the Hermite functions $\Phi_\alpha$ constitute a basis, any $\psi\in
L^2(\mathbb{R}^n)$ can be expanded as
\begin{equation}\label{eq:hermite-exp} \psi(x) = \sum_{\alpha} c_\alpha  \Phi_\alpha(x), \quad \|\psi\|^2 =
\sum_{\alpha} |c_\alpha|^2, \end{equation} where the sum is to be taken over all
multi-indices $\alpha\in\Ind_n$. Now, let $\beta\in \Ind_n$ be arbitrary. By
using Eqn.~(\ref{eq:ladder2}) in each spatial direction we compute the
action of $(\mathbf{a}^\dag)^\beta$ on $\psi$:
\begin{eqnarray}
  (\mathbf{a}^\dag)^\beta\psi &=& (a_1^\dag)^{\beta_1} \cdots (a_n^\dag)^{\beta_n}\sum_\alpha
  c_\alpha \Phi_\alpha  \nonumber \\ 
  &=& \sum_{\alpha} c_{\alpha} \prod_{k=1}^n
  \frac{(\alpha_k+\beta_k)!^{1/2}}{\alpha_k!^{1/2}}
  \Phi_{\alpha+\beta} \nonumber \\&=& \sum_\alpha c_\alpha P_\beta(\alpha)^{1/2}
  \Phi_{\alpha+\beta}. 
\end{eqnarray}
Similarly, by using Eqn.~(\ref{eq:ladder}) we obtain
\begin{eqnarray}
  \mathbf{a}^\beta\psi &=& a_1^{\beta_1} \cdots a_n^{\beta_n}\sum_\alpha
  c_\alpha \Phi_\alpha  \nonumber \\ 
  &=& \sum_{\alpha} c_{\alpha+\beta} \prod_{k=1}^n
  \frac{(\alpha_k+\beta_k)!^{1/2}}{\alpha_k!^{1/2}}
  \Phi_\alpha \nonumber \\&=& \sum_\alpha c_{\alpha+\beta} P_\beta(\alpha)^{1/2} \Phi_\alpha,
\end{eqnarray}
Computing the square norm gives
\begin{equation} \| (\mathbf{a}^\dag)^\beta\psi \|^2 = \sum_{\alpha} P_\beta(\alpha) |c_{\alpha}|^2.  \label{eq:norm1}\end{equation}
and
\begin{equation} \| \mathbf{a}^\beta\psi \|^2 = \sum_{\alpha} P_\beta(\alpha)
|c_{\alpha+\beta}|^2. \label{eq:norm2} \end{equation}
The polynomial $P_\beta(\alpha)>0$ for all $\beta,\alpha\in\Ind_n$, and
$P_\beta(\alpha+\alpha')>P_\beta(\alpha)$ for all non-zero
multi-indices $\alpha'\neq 0$. Therefore, if
$(\mathbf{a}^\dag)^\beta\psi\in L^2(\mathbb{R}^n)$ then $\mathbf{a}^\beta\psi\in
L^2(\mathbb{R}^n)$. However, the converse is not true for $n>1$ dimensions, as the norm in
Eqn.~(\ref{eq:norm2}) is independent of infinitely many coefficients
$c_\alpha$, while Eqn.~(\ref{eq:norm1}) is not. (This should be
contrasted with the one-dimensional case, where $a\psi\in
L^2(\mathbb{R})$ was \emph{equivalent} to $a^\dag\psi\in
L^2(\mathbb{R})$.) On the other hand, as in the $n=1$ case, the condition $a_k^\dag\psi\in L^2(\mathbb{R}^n)$ is equivalent to
the conditions $x_k\psi\in L^2(\mathbb{R}^n)$ and $\partial_k\psi\in
L^2(\mathbb{R}^n)$. 

We are in position to formulate a straightforward generalization of
Lemma \ref{lemma:approx1}. The proof is easy, so we omit it.
\begin{lemma}[General Hermite expansions]
\label{lemma:props-of-hermite-exp}
Let $\psi\in L^2(\mathbb{R}^n)$, with coefficients $c_\alpha$ as in
Eqn.~(\ref{eq:hermite-exp}), and let $\beta\in\Ind_n$ be
arbitrary. Assume $(\mathbf{a}^\dag)^\beta\psi\in L^2(\mathbb{R}^n)$. Then
$(\mathbf{a}^\dag)^{\beta'}\psi\in L^2(\mathbb{R}^n)$ and
$\mathbf{a}^{\beta'}\psi\in L^2(\mathbb{R}^n)$ for all $\beta'\leq\beta$. Moreover,
the following points are equivalent:
\begin{enumerate}
\item \label{thm:point-i}
  $(\mathbf{a}^\dag)^\beta\psi\in L^2(\mathbb{R}^n)$
\item \label{thm:point-ii}
  For all multi-indices $\gamma\leq\beta$,
  $x^\gamma\partial^{\beta-\gamma}\psi\in L^2(\mathbb{R}^n)$.
\item \label{thm:point-iii}
  $\sum_\alpha \alpha^\beta |c_\alpha|^2 < +\infty$
\end{enumerate}
\end{lemma}

We observe, that as we obtained for $n=1$, condition
\ref{thm:point-ii} is a combined decay and smoothness condition on
$\psi$, and that this can be expressed as a decay-condition on the
coefficients of $\psi$ in the Hermite basis by \ref{thm:point-iii}.

Exponential decay of $\psi\in L^2(\mathbb{R}^n)$ as
$\|x\|\rightarrow\infty$ implies that that $x^\gamma\psi\in
L^2(\mathbb{R}^{n})$ for all $\gamma\in\Ind_n$. We now
generalize Lemma \ref{lemma:exp1} to the $n$-dimensional case. 

\begin{lemma}[Exponentially decaying functions]
\label{thm:exp-decay}
Assume that $\psi\in L^2(\mathbb{R}^n)$ is such that for all
$\gamma\in\Ind_n$, $x^\gamma\psi\in L^2(\mathbb{R}^n)$. Then, a
sufficient criterion for $(\mathbf{a}^\dag)^\beta\psi\in
L^2(\mathbb{R}^n)$ is $\partial^{\beta}\psi\in L^2(\mathbb{R}^n)$.  Moreover, for
all $\mu\leq\beta$, we have $x^\gamma\partial^{\beta - \mu}
\psi \in L^2(\mathbb{R}^n)$ for all $\gamma\in\Ind_n$ such that
$\gamma_k=0$ whenever $\mu_k=0$, i.e., the partial derivatives of
lower order than $\beta$ decay exponentially in the directions where
the differentiation order is lower.
\end{lemma}

\emph{Proof:} The proof is a straightforward application of the $n=1$
case in an inductive proof, together with the following elementary
fact concerning weak derivatives: If $1\leq j < k \leq n$, and if
$x_j\psi(x)$ and $\partial_k\psi(x)$ are in $L^2(\mathbb{R}^n)$, then,
by the product rule, $\partial_k(x_j\psi(x)) = x_j(\partial_k\psi(x))
\in L^2(\mathbb{R}^n)$. Notice, that Lemma~\ref{lemma:exp1} trivially
generalizes to a single index in $n$ dimensions, i.e., to $\beta =
\beta_ke_k$, since the integration by parts formula used is valid in
$\mathbb{R}^n$ as well. Similarly, the present Lemma is valid in $n-1$
dimensions if it holds in $n$ dimensions, as it must be valid for
$\bar{\beta} = (0,\beta_2,\cdots,\beta_n)$.

Assume that our statement holds for $n-1$ dimensions. We must prove
that it then holds in $n$ dimensions. Assume then, that
$\partial^\beta\psi\in L^2(\mathbb{R}^n)$. Let $\phi =
\partial^{\bar{\beta}}\psi \in L^2$. Moreover,
$\partial_1^{\beta_1}\phi\in L^2$. Since $\psi$ is exponentially
decaying, and by the product rule, $x_1^{\gamma_1}\phi\in L^2$ for
all $\gamma_1\geq 0$. By Lemma~\ref{lemma:exp1},
$x_1^{\gamma_1}\partial_1^{\beta_1-\mu_1}\phi\in L^2$ for all
$\gamma_1$ and $0<\mu_1\leq \beta_1$. Thus,
$x_1^{\gamma_1}\partial^{\beta-e_1\mu_1}\psi\in L^2$. Thus, the
result holds as long as $\mu = e_1\mu_1$; or equivalently $\mu =
e_k\mu_k$ for any $k$. To apply induction, let $\chi =
x_1^{\gamma_1}\partial_1^{\beta_1-\mu_1}\psi \in L^2$. Note that
$\partial^{\bar{\beta}}\chi\in L^2$ and $x^{\bar{\gamma}}\chi\in
L^2$ for all $\bar{\gamma}=(0,\gamma_2,\cdots,\gamma_n)$. But by the
induction hypothesis,
$x^{\bar{\gamma}}\partial^{\bar{\beta}-\bar{\mu}}\chi\in L^2$ for
all $\bar{\mu}\leq\bar{\beta}$ and all $\bar{\gamma}$ such that
$\bar{\gamma}_k=0$ if $\bar{\mu}_k=0$. This yields, using the product
rule, that $x^\gamma\partial^{\beta-\mu}\psi\in L^2$ for all
$\mu\leq\beta$ and all $\gamma$ such that $\gamma_k=0$ if $\mu_k=0$,
which was the hypothesis for $n$ dimensions, and the proof is
complete. Notice, that we have proved that
$(\mathbf{a}^\dag)^\beta\psi\in L^2$ as a by-product.
\eofproof

In order to generate a simple and useful result for approximation in
$n$ dimensions, we consider the case where $\psi$ decays
exponentially, and $\psi\in H^k(\mathbb{R}^n)$, i.e.,
$\partial^\beta\psi(x)\in L^2(\mathbb{R}^n)$ for all
$\beta\in\Ind_n$ with $|\beta|=k$.  In this case, we may also
generalize Eqn.~(\ref{eq:order-1d}). For this, we consider the
shell-weight $p(r)$ defined by
\begin{equation}
p(r) \equiv \sum_{\alpha\in\Ind_n,\;|\alpha|=r} |c_\alpha|^2,
\label{eq:shell-prob}
\end{equation}
where $c_\alpha = \langle \Phi_\alpha, \psi \rangle$. Then,
$\|\psi\|^2 = \sum_{r=0}^\infty p(r)$. Moreover, if $P$ projects onto
the shell-truncated Hilbert space $\Proj_R(\mathbb{R}^n)$, then 
\begin{equation} \|P\psi\|^2 = \sum_{r = 0}^R p(r). \end{equation}

\begin{theorem}[Approximation in $n$ dimensions]
Let $\psi \in L^2(\mathbb{R}^n)$ be exponentially decaying and given by
\begin{equation} \psi(x) = \sum_\alpha c_\alpha \Phi_\alpha(x). \end{equation}
Then $\psi\in H^k(\mathbb{R}^n)$, $k\geq 0$, if and only if
\begin{equation} \sum_\alpha |\alpha|^k |c_\alpha|^2 = \sum_{r=0}^\infty r^k p(r)  <
+\infty. \label{eq:multisum1} \end{equation}
The latter implies that 
\begin{equation} p(r) =
  o(r^{-(k+1)}).  \end{equation} Moreover,
for the shell-truncated Hilbert space $\Proj_R$, the approximation
error is given by
\begin{equation} \| (1-P)\psi \| = \left( \sum_{r=R+1}^\infty
  p(r)\right)^{1/2}. 
\end{equation}
\end{theorem}

\emph{Proof:} The only non-trivial part of the proof concerns
Eqn.~(\ref{eq:multisum1}). Since $\psi$ is exponentially decaying and since
  $\psi\in H^k$ if and only if $\partial^\beta\psi\in L^2$ for all
  $\beta$, $|\beta|\leq k$, we know that $\sum_\alpha
  \alpha^\beta|c_\alpha|^2 < +\infty$ for all $\beta$, $|\beta|\leq
  k$. Since $|\alpha|^k$ is a polynomial of order $k$ with terms of
  type $a_\beta \alpha^\beta$, $a_\beta\geq 0$ and $|\beta| = k$, we
  have
\begin{equation} \sum_\alpha |\alpha|^k |c_\alpha|^2 = \sum_{\beta,\;|\beta|=k}
a_\beta \sum_{\alpha} \alpha^\beta |c_\alpha|^2 < +\infty. \end{equation}
On the other hand, since $a_\beta\geq 0$ and the sum over $\beta$ has
finitely many terms, $\sum_\alpha|\alpha|^k|c_\alpha|<+\infty$ implies
$\sum_\alpha \alpha^\beta|c_\alpha|^2<+\infty$ for all $\beta$,
$|\beta| = k$, and thus $\psi\in H^k$ since $\psi$ was exponentially
decaying. 
\eofproof

\end{document}